\documentclass[useAMS,usenatbib]{mn2e}
\usepackage{subfiles}
\usepackage[english]{babel}
\usepackage{natbib}
\usepackage{graphics}
\usepackage{graphicx}
\usepackage{color}
\usepackage{hyperref}
\usepackage{amssymb}
\def\simlt{\lower.5ex\hbox{$\; \buildrel < \over \sim \;$}}
\def\simgt{\lower.5ex\hbox{$\; \buildrel > \over \sim \;$}}

\newcommand{\kms}{\,\mathrm{km \, s^{-1}}}
\newcommand{\msol}{\,\mathrm{M_{\sun}}}

\title[Galaxy structure with strong gravitational lensing]
{Galaxy structure with strong gravitational lensing: \\
decomposing the internal mass distribution of massive elliptical galaxies}
\author[Nightingale et al.]
{\parbox{\textwidth}{James. W. Nightingale$^{1}$\thanks{e-mail: james.w.nightingale@durham.ac.uk},
Richard J.\ Massey$^{1}$,
David R.\ Harvey$^{2}$, 
Andrew P.\ Cooper$^{1,3}$,
Amy Etherington$^{1}$,
Sut-Ieng Tam$^{1}$ \&
Richard G. Hayes$^{1}$ \\
}\\
$^{1}$Centre for Extragalactic Astronomy, Department of Physics, Durham University, South Road, Durham, DH1 3LE, UK\\
$^{2}$Laboratoire d'Astrophysique, EPFL, Observatoire de Sauverny, 1290 Versoix, Switzerland\\
$^{3}$Institute of Astronomy and Department of Physics, National Tsing Hua University, Hsinchu 30013, Taiwan 
}

\begin{document}

\bibliographystyle{mn2e}
\bibpunct{(}{)}{;}{a}{}{;}
\date{\today}
\pagerange{\pageref{firstpage}--\pageref{lastpage}} 
\pubyear{2018}
\maketitle
\label{firstpage}

\begin{abstract}

We investigate how strong gravitational lensing can test contemporary models of massive elliptical (ME) galaxy formation, by combining a traditional decomposition of their visible stellar distribution with a lensing analysis of their mass distribution. As a proof of concept, we study a sample of three ME lenses, observing that all are composed of two distinct baryonic structures, a `red' central bulge surrounded by an extended envelope of stellar material. Whilst these two components look photometrically similar, their distinct lensing effects permit a clean decomposition of their mass structure. This allows us to infer two key pieces of information about each lens galaxy: (i) the stellar mass distribution (without invoking stellar populations models) and (ii) the inner dark matter halo mass. We argue that these two measurements are crucial to testing models of ME formation, as the stellar mass profile provides a diagnostic of baryonic accretion and feedback whilst the dark matter mass places each galaxy in the context of LCDM large scale structure formation. We also detect large rotational offsets between the two stellar components and a lopsidedness in their outer mass distributions, which hold further information on the evolution of each ME. Finally, we discuss how this approach can be extended to galaxies of all Hubble types and what implication our results have for studies of strong gravitational lensing.

\end{abstract}


\section{INTRODUCTION}

The work of \citet{Hubble1926} famously classified galaxies into three groups: ellipticals, spirals and irregulars. Today, using samples of hundreds of thousands of galaxies, these classifications have been broadly established to hold across all galaxies, in the local Universe \citep{Hoyos2011, Vika2013, Vulcani2014} and at high redshift \citep{Buitrago2008, Chevance2012, Bruce2012, VanDerWel2012}. The bulk structure of a galaxy can be quantified by its one-dimensional projected surface brightness profile. The Sersic function, $\mu({R}) \propto R^{\frac{1}{n}}$,  \citep{Vaucouleurs1948, Sersic1968} has proven extremely useful for this purpose across the entire Hubble sequence. The bulk of the light in a typical elliptical galaxy is well described by $n\sim4$ and an exponential disk structure by $n\sim1$. Although most galaxies can be labelled with a specific morphology on the Hubble diagram without much ambiguity, they can also exhibit sub-dominant structures with other morphologies \citep[see][]{Graham2011}. For example, spirals have bulges \citep{Thomas2007, MacArthur2009a}, ellipticals have disks \citep{Oosterloo2002, Bluck2015} and irregulars may show signs of both bulges and disks \citep{Bentabol2016}. This has motivated descriptions that use multiple Sersic profiles \citep{Lackner2012, Bruce2014, Vika2014, Kennedy2015} to decompose galaxies into their constituent physical structures.

Fitting the light distribution of a galaxy with a superposition of light profiles is a challenging and highly degenerate problem  \citep[e.g.][]{Haussler2013}. For example, the extended wings of a central bulge can be difficult to separate from structures further out (e.g. a disk), as they blend with one another and the background sky emission. Bulges, disks and bars may all appear as compact central structures in a galaxy, which are hard to decouple from one another \citep{Kormendy2004}. Absorption of light by dust can also impact the fit, leading to more ambiguities in the interpretation of a galaxy's structure \citep{Rest2001, Lauer2005}.  

Using integral field spectroscopy (IFS), the SAURON \citep{Bacon2001} and ATLAS3D \citep{Cappellari2011} surveys demonstrated the importance, when trying to infer a galaxy's structure, of having data sensitive to its mass. For example, \citet{Emsellem2004, Emsellem2007, Emsellem2011} found that $\sim 90\%$ of galaxies in a volume-limited sample showed some level of ordered rotation in their kinematics despite showing no signs of a disk in their light \citep{Krajnovic2011, Krajnovic2013}. IFS data also reveal that biases in the inferred galaxy structure may arise due to the 3D inclination \citep{Devour2017} and the triaxiality \citep{Mendez-Abreu2010, Mendez-Abreu2010a} of the different structures within galaxies, which limits inferences based on their 2D projected morphologies \citep{Cappellari2008, Emsellem2011, Weijmans2014}.

In this work, we propose that strong gravitational lensing can both mitigate the degeneracies that arise when fitting a galaxy's 2D light distribution and provide key insight on its underlying physical mass structure. Strong lensing is the deflection of light from a background source around a foreground lens galaxy, giving rise to multiple images of the source with characteristic distortions (see \citet{Kochanek2004a} for an overview). These distortions encode information on the foreground lens's mass distribution and make it easier to separate the different galaxy components in comparison to using photometry alone. Conveniently, lensing data can be extracted from the same CCD imaging as the photometry, in contrast to IFS data which must be obtained separately: at significant cost in observing time, and typically at much coarser spatial resolution.

For this proof-of-concept study, we use our open-source lens modeling software {\tt PyAutoLens}\footnote{https://github.com/Jammy2211/PyAutoLens} \citep{Nightingale2015, Nightingale2018} to fit the projected optical luminosity distribution of three isolated massive elliptical (ME) galaxies with velocity dispersions of $160$-$250\;\kms$, whilst {\em simultaneously} constraining their underlying mass structure via a strong lensing analysis. We recover the distribution of both light and mass in projection along the line of sight, enabling these two inferences of galaxy structure to be compared directly and circumventing ambiguities due to dust. Crucially, this approach allows us to confirm whether or not features apparent in the light distribution correspond to genuine physical structures.

This question is central to tests of ME galaxy formation in the LCDM cosmology, which predicts that they assemble their stellar mass both by dissipative star formation and by mergers with other galaxies. Each of these processes should give rise to physically distinct components in the phase-space structure of MEs that may have observable signatures in their stellar mass surface density profiles. In particular, active galactic nuclei (AGN) activity at the centre of haloes more massive than $\sim10^{11}\msol$ is thought to act as a `thermostat' that shuts down in-situ growth of their central galaxy by suppressing radiative cooling of fresh cold gas from a hot circumgalactic reservoir \citep{Croton2006, Bower2006, Bower2017}. The progenitors of present-day MEs form rapidly in overdense regions at $z\simgt 2$--$4$, hence the stars they form in situ (i.e. before AGN suppression) are characterised by high metallicity and high phase space density \citep{Baugh1996, Hopkins2006, DeLucia2005}. By the present day these stellar populations have old ages and red colours.

Although in situ star formation is suppressed in MEs, DM mass growth continues in an approximately self-similar fashion \citep[e.g.][]{Guo2008a}. The most massive halos at the present day, which are much more massive than the `threshold' mass for AGN suppression, coalesce relatively recently. Their immediate progenitor halos are themselves likely to be above the threshold mass, and hence also to host gas-poor galaxies with `red and dead' metal-rich stellar populations. This leads to a picture in which the bulk of stellar mass in MEs is assembled through dissipationless mergers of several equally important fragments. These fragments have a broad range of DM halo masses but a narrow range of stellar masses \citep{DeLucia2007}. 

The structure of the resulting MEs (represented to lowest order by the scale and shape of their surface brightness profile) is therefore dominated by the `initial conditions' and gravitational dynamics of these mergers, in which DM plays a dominant role \citep[e.g.][]{Cole2000,Boylan-Kolchin2006,Naab2009,Hilz2013,Laporte2013}. Broadly speaking, forward models of ME formation predict a composite structure arising from the superposition of phase-mixed stellar debris from each progenitor, each with a characteristic profile and scale \citep[similar to the assembly of stellar halos around less massive galaxies described by][]{Amorisco2017}. For typical ME assembly histories, the dominant accreted components are expected to have similar amplitude, shape and radial extent \citep[e.g.][]{Cooper2013, Cooper2015}. Consequently, the aggregate stellar density profile is (on average) unlikely to show strong inflections that correspond to transitions between regions of the galaxy dominated by debris from different progenitors. Indeed, observations (e.g. \citet{Kormendy2009}) show that to good approximation, the overall structure may be described as a single component with Sersic $n\simgt4$ \citep{Schombert2015a}.

The strongest observational features are therefore most likely to arise between the more extended aggregate accreted component and the centrally concentrated component, corresponding to an in-situ, high redshift `nugget' (with much higher phase space density as the result of dissipative formation). Sufficiently deep imaging of local MEs observe this \citep{Huang2013a,Huang2017,Spavone2017}, including components with different ellipticities, orientations (isophotal twists) \citep{Oh2016, Oh2017b} and centers \citep{Goullaud2018}. Spectroscopic studies of MEs are also consistent with this picture, whereby the stellar ages and alpha-element abundances of their central regions are both observed to increase with velocity dispersion \citep{Greene2012, Greene2013, Greene2015a}, trends which appear washed out at larger radii where stars are anticipated to have been more recently accreted\footnote{These expectations for accretion on to MEs are significantly different to those for Milky Way-like galaxies, which correspond to the regime of in situ dominated growth in lower-mass halos largely unaffected by AGN feedback. In that regime, the analogous accreted stellar component (the stellar halo) is expected to be dominated by only one or two accretion events of high mass ratio, and to comprise only a small fraction of the total mass. Stellar halo progenitors arrive on more circular orbits (giving rise to streams and other unrelaxed structures) and consist of stellar populations very different to those of the central galaxy \citep{Bullock2005,Cooper2010,Amorisco2017}.}. 

Observations such as these have been interpreted as evidence for a model in which an `outer envelope' arises from prolonged and significant accretion onto a high-density stellar core. This model, often referred to as `two phase' assembly \citep[e.g.][]{Oser2010a}, is broadly equivalent to the predictions of the LCDM models described above in the ME regime. Examples of massive compact ellipticals, thought to correspond to these dissipative cores before significant accretion, have been observed at high redshift but are extremely rare at the present day \citep[e.g.][]{Trujillo2006, Szomoru2011a,Oldham2016}. Overall, observations appear to be consistent with the evolution of the galaxy population as a whole predicted by LCDM models, once selection effects are taken into account \citep{Laporte2013,Xie2015,Furlong2017,Roy2018}. Furthermore, indirect tracers of dark matter mass (e.g.\ environment or size) detect correlations consistent with the above picture in populations of elliptical galaxies \citep{Lani2013, Sonnenfeld2015, Huang2018}.

Nevertheless, decomposing MEs into multiple components remains very challenging, because their two (or more) superposed elliptical components (with similar stellar populations) are often well described by a single elliptical profile. This makes them the ideal case-study for investigating how strong lensing can aid the decomposition of a galaxy's structure. We will demonstrate that strong lens modeling: (i) allows us to confirm that the structures we see visualally correspond to genuine mass components; (ii) provides direct access to the stellar mass distribution of each component (without stellar population modeling) and; (iii) infers the central ($\approx 10$ kpc) dark matter halo mass of each galaxy. With a sample of just $3$ objects, our conclusions primarily focus on discussing the utility of this method. In the future, existing samples of hundreds of lenses will enable a direct test of LCDM expectations for the assembly of MEs, in particular the fundamental relationship between halo mass, stellar mass distribution and galaxy history. Future samples of 100000 strong lenses \citep{Collett2015} will allow such an analysis to be generalised over the entire Hubble sequence.


This paper is structured as follows.
In \S\ref{Data}, we describe our sample of three ME galaxies selected for detailed study.
In \S\ref{Method}, we describe the {\tt PyAutoLens} method for simultaneous photometric and strong lensing analysis.
In \S\ref{Results}, we investigate the mass structure of our lenses.
In \S\ref{Discussion}, we discuss the implications of our measurements, and we give a summary in \S\ref{Summary}.
{We assume a Planck 2015 cosmology throughout \citep{PlanckCollaboration2015a}}.

\section{Data Reduction and Lens Sample}\label{Data}

\subsection{HST data reduction}\label{Method_dr}

We adopt a modified data reduction pipeline that uses a combination of in-house tools and those from the standard Space Telescope Science Institute library. We first correct images for charge transfer inefficiency using the {\tt arCTIc} software \citep{Massey2010,Massey2014}. We then bias subtract, flat field and co-add multiple exposures using the {\tt calacs} and {\tt astrodrizzle} packages to create a final data product. Following \cite{Rhodes2007}, we combine the images using a square drizzle kernel and a pixel fraction of 0.8, to a final pixel scale of $0.03"$. To determine the Point Spread Function (PSF) of this final image, we measure the focus of the telescope during each exposure, via the quadrupole moments of stars throughout that exposure \citep[c.f.][]{Harvey2015}, then coadd (appropriately rotated) {\tt TinyTim} models of each exposure's PSF \citep{Krist2004}. We store a single, 2D model of the combined PSF at the location of the lens, with a matching pixel scale of 0.03$\arcsec$, with the peak of the PSF the centre of a pixel.

This procedure produces a sky-subtracted, stacked image of each lens, which is used for the analysis. The variance in each pixel is estimated as $\sqrt{\sigma^2_{\mathrm{BG}} + d_{\mathrm{j}}}$, where $\sigma^2_{\mathrm{BG}}$ is the estimated variance in the background and $d_{\mathrm{j}}$ is the total counts detected in pixel $j$ (accounting for the variation in exposure time due to dithering). The variances and background sky level are included as part of the modeling procedure (see N18).

\subsection{Lens Sample}

\begin{figure*}
\centering
\includegraphics[width=0.325\textwidth]{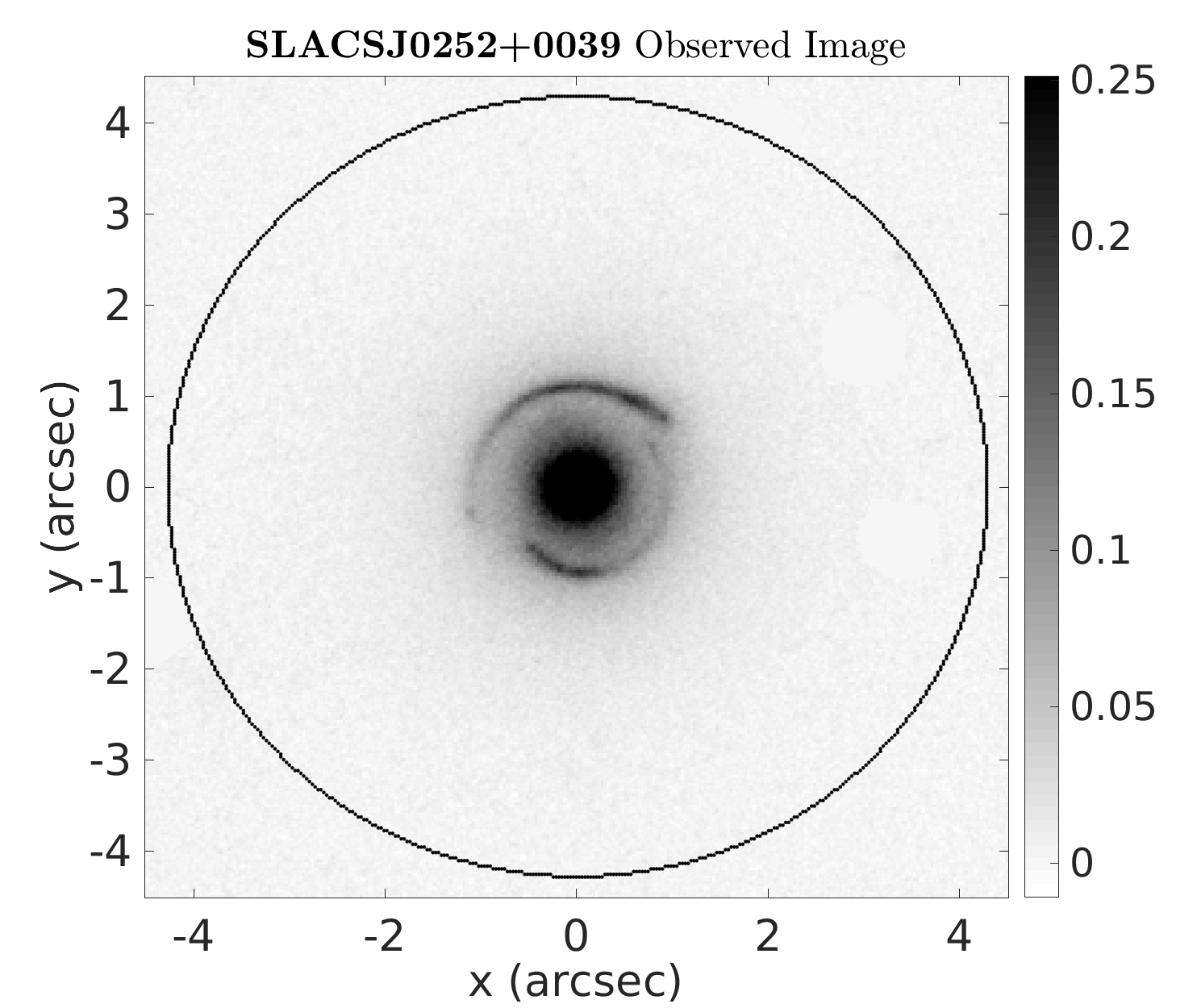}
\includegraphics[width=0.325\textwidth]{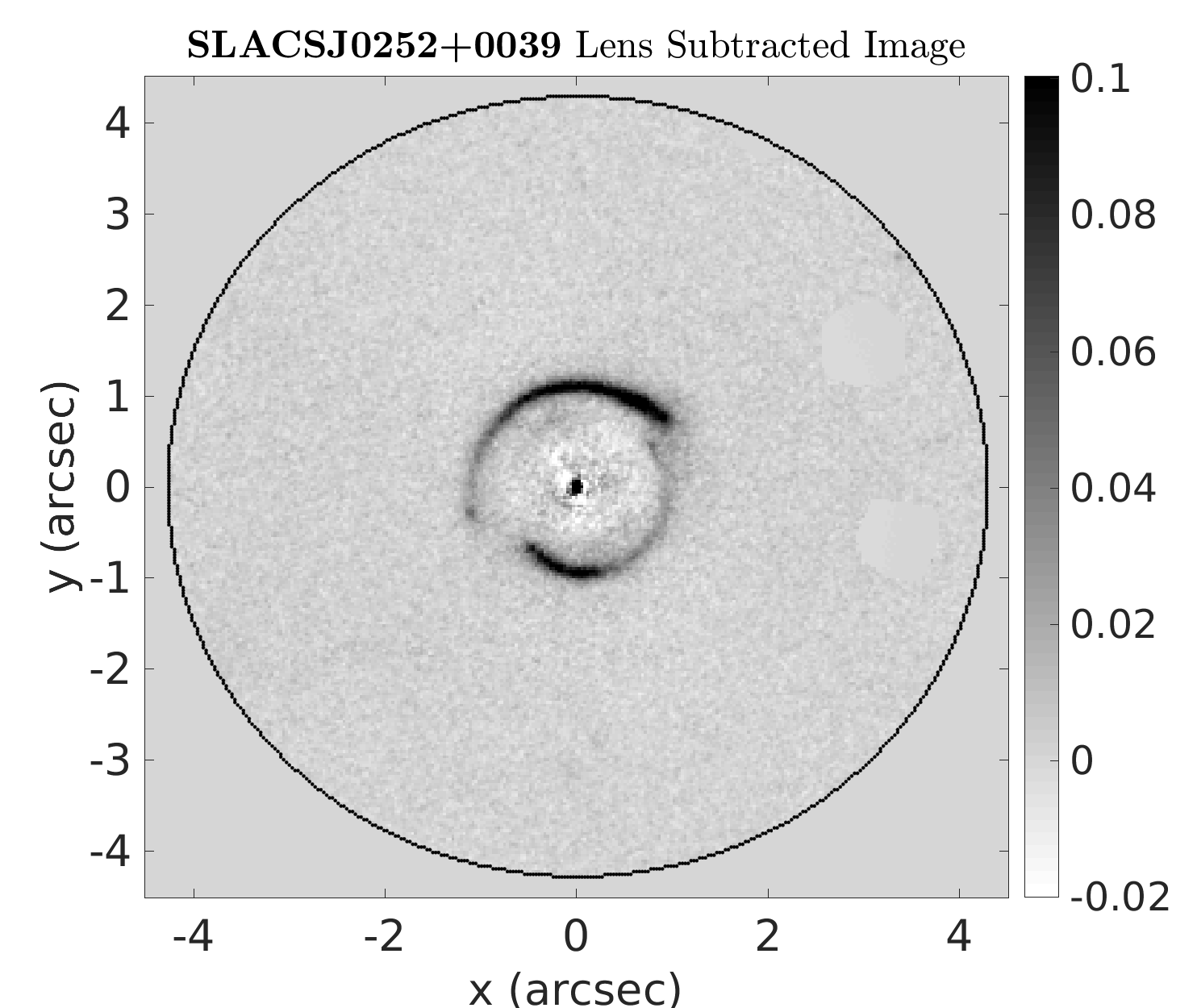}
\includegraphics[width=0.325\textwidth]{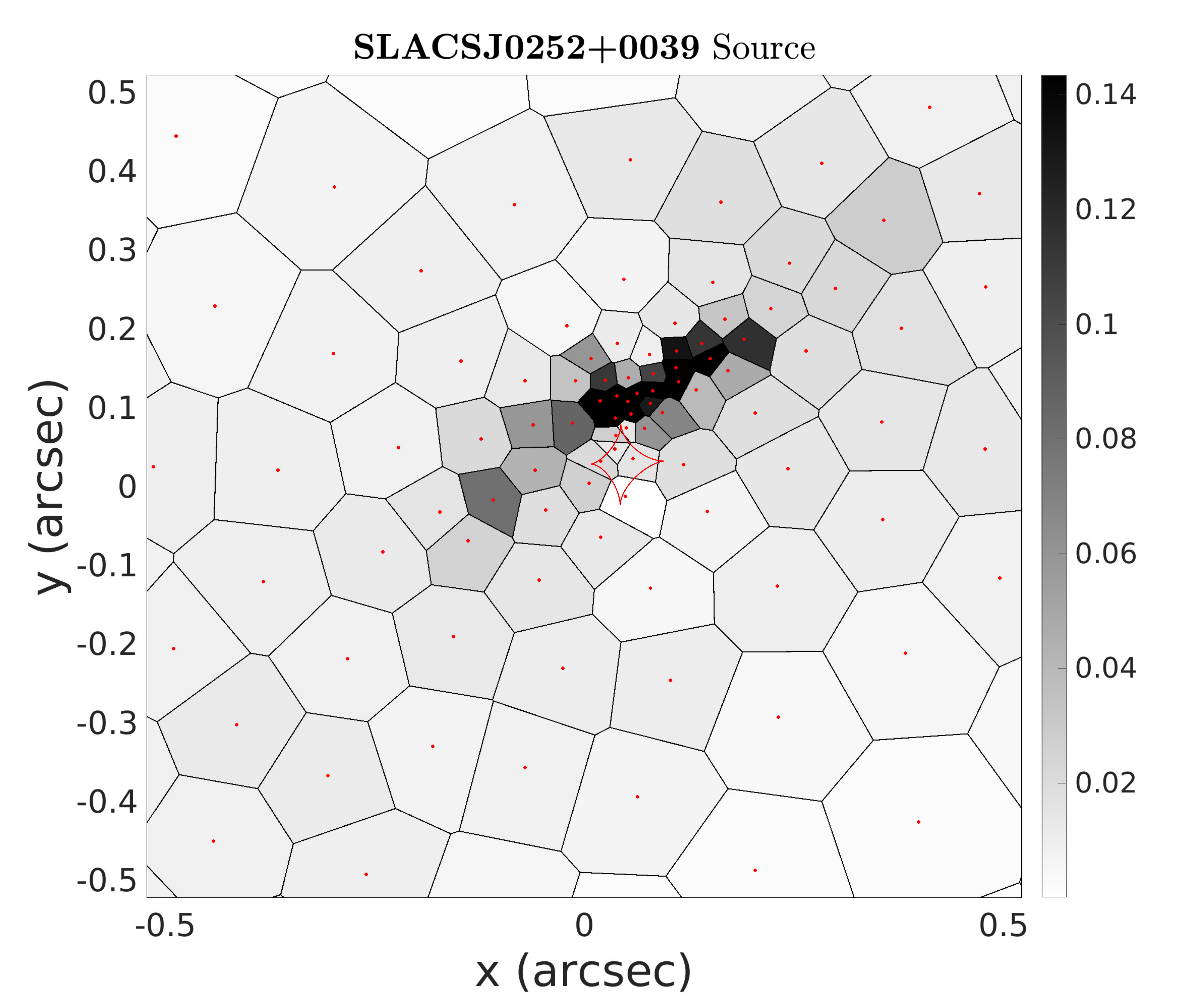}
\includegraphics[width=0.325\textwidth]{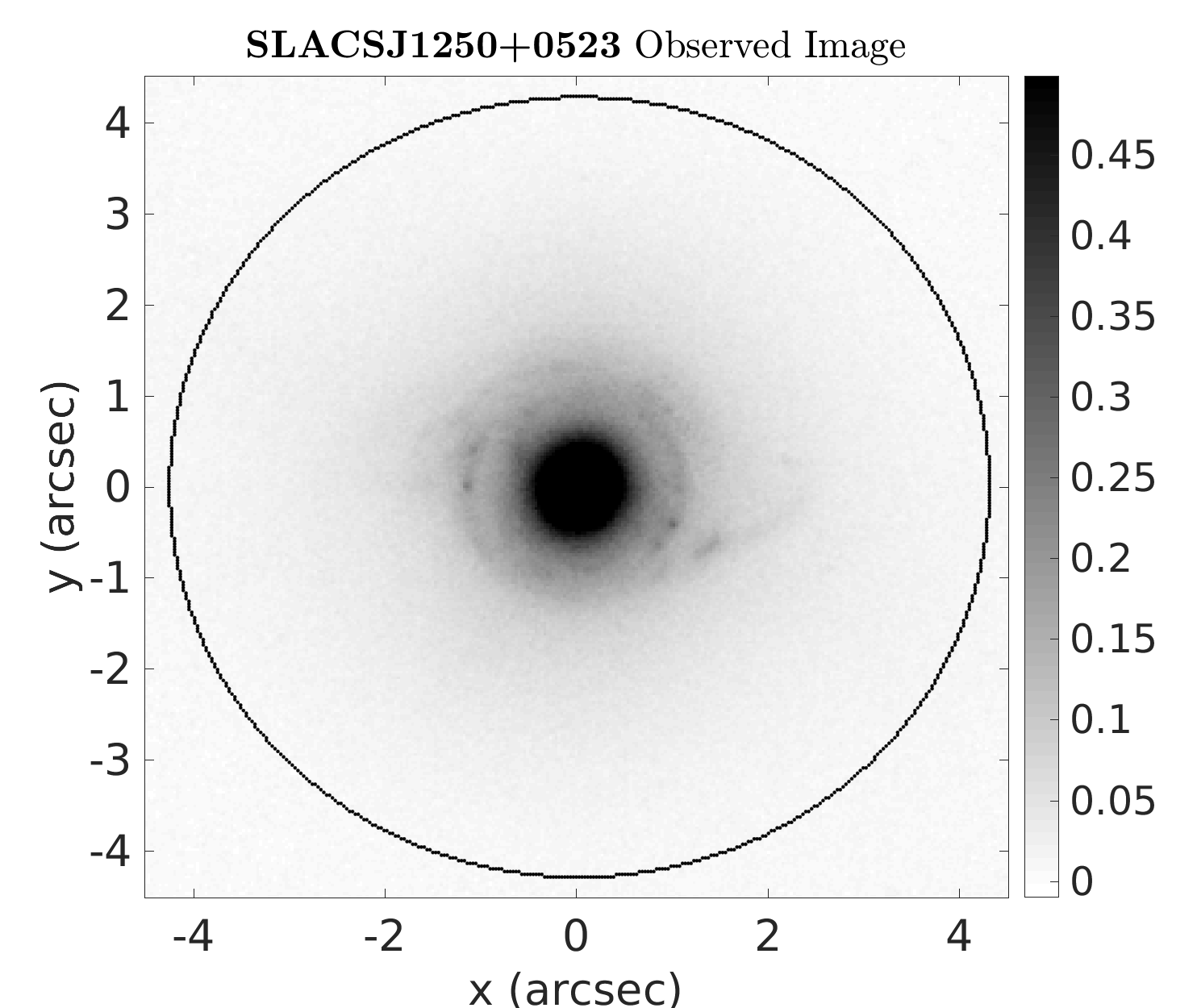}
\includegraphics[width=0.325\textwidth]{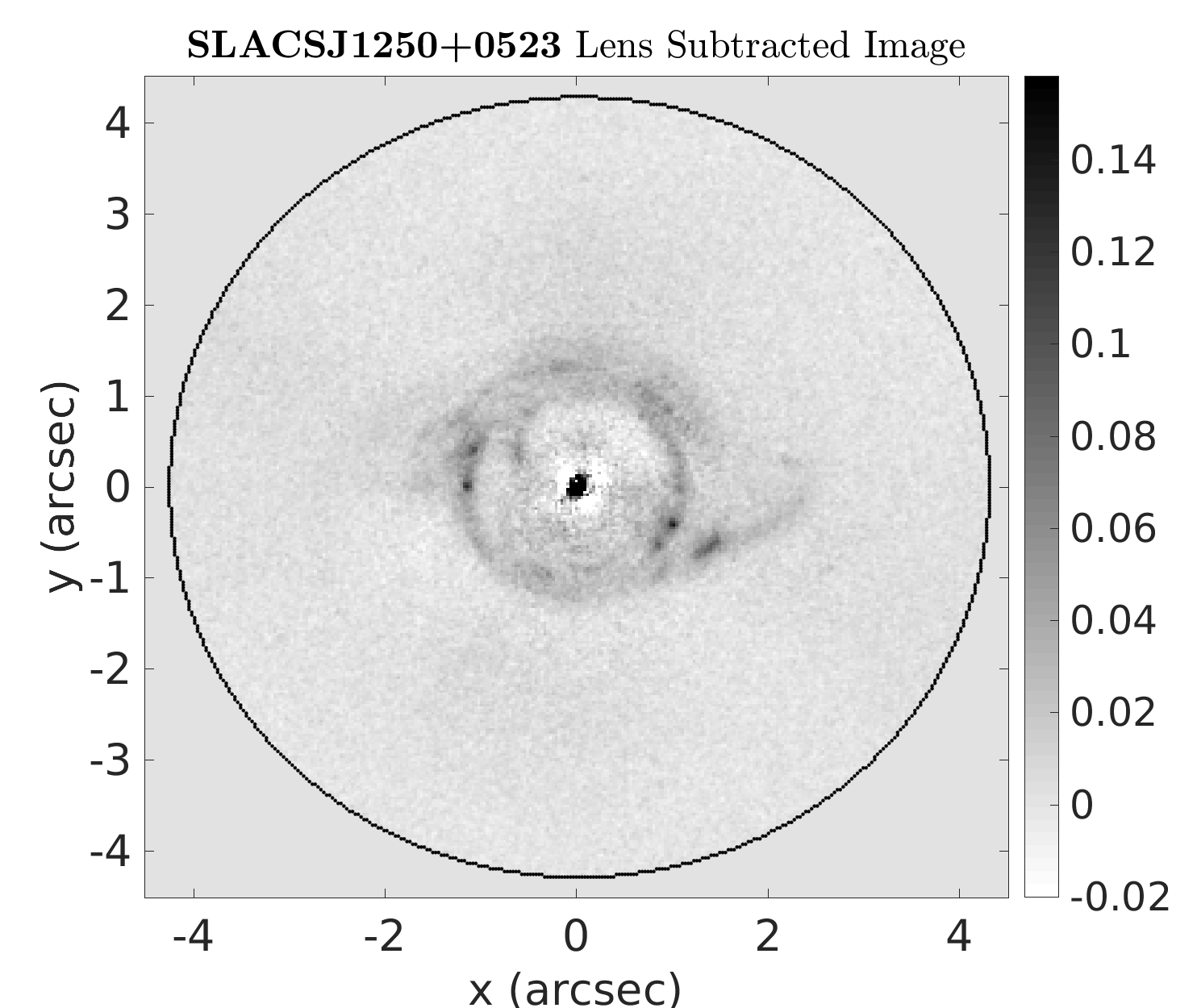}
\includegraphics[width=0.325\textwidth]{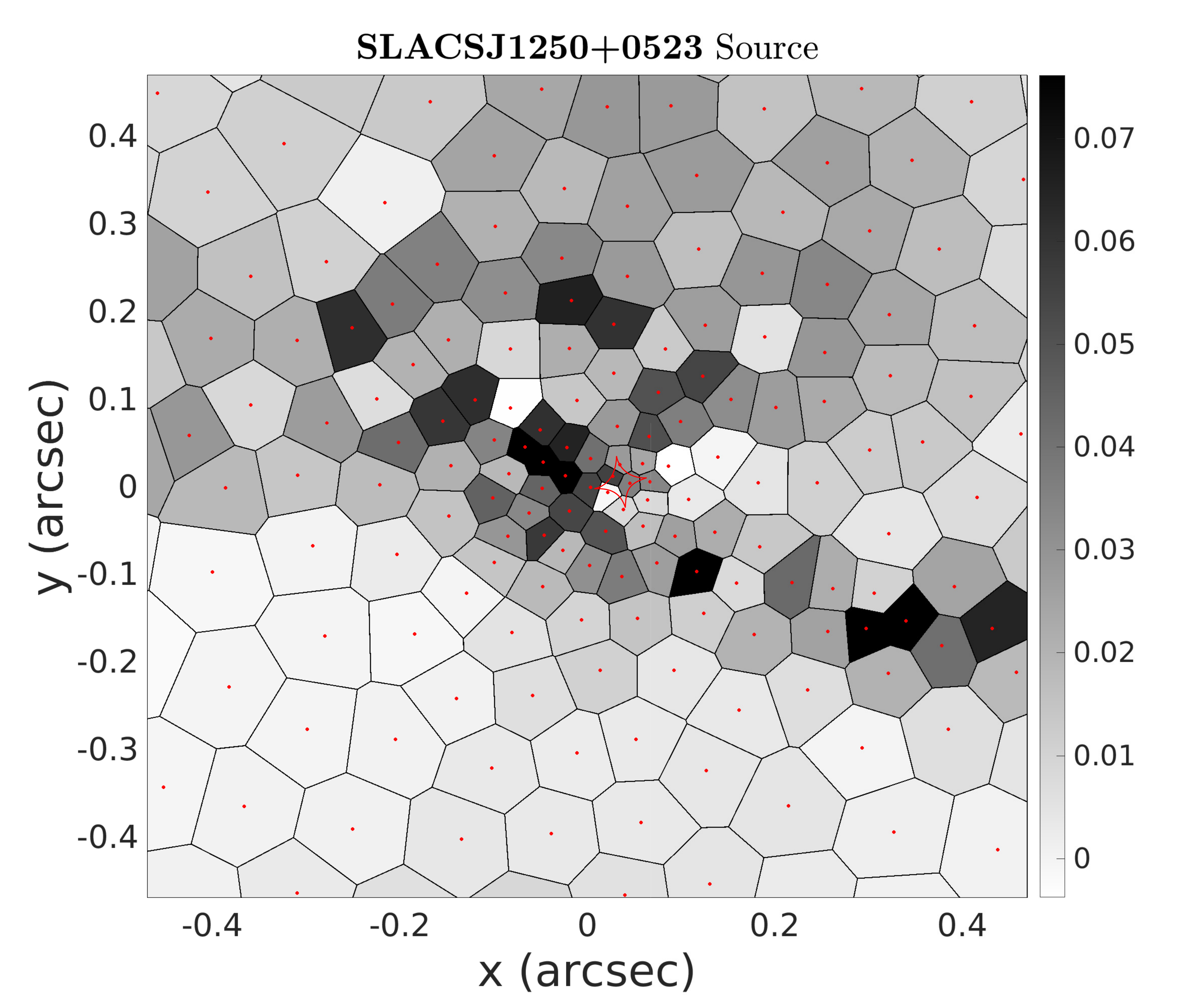}
\includegraphics[width=0.325\textwidth]{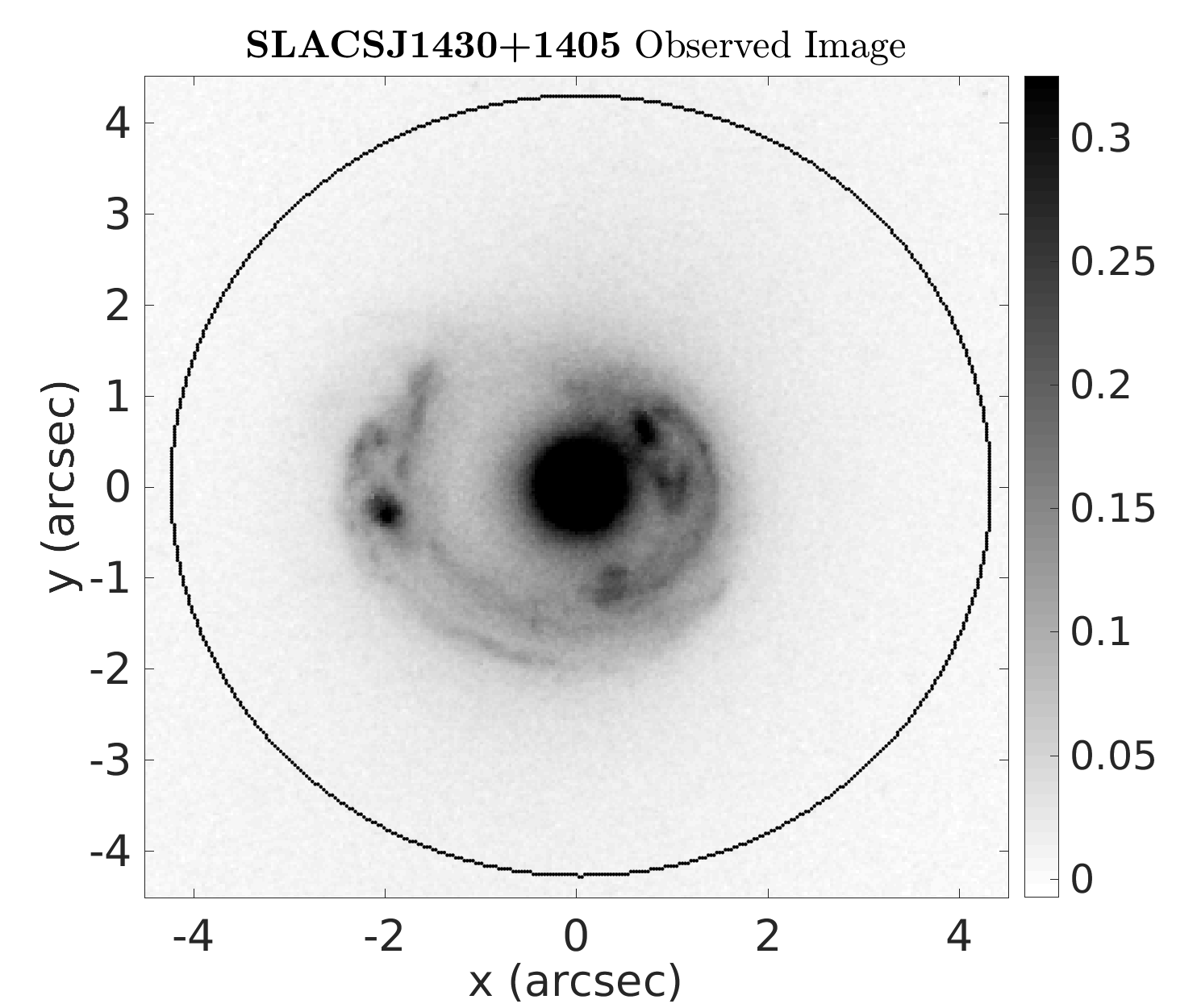}
\includegraphics[width=0.325\textwidth]{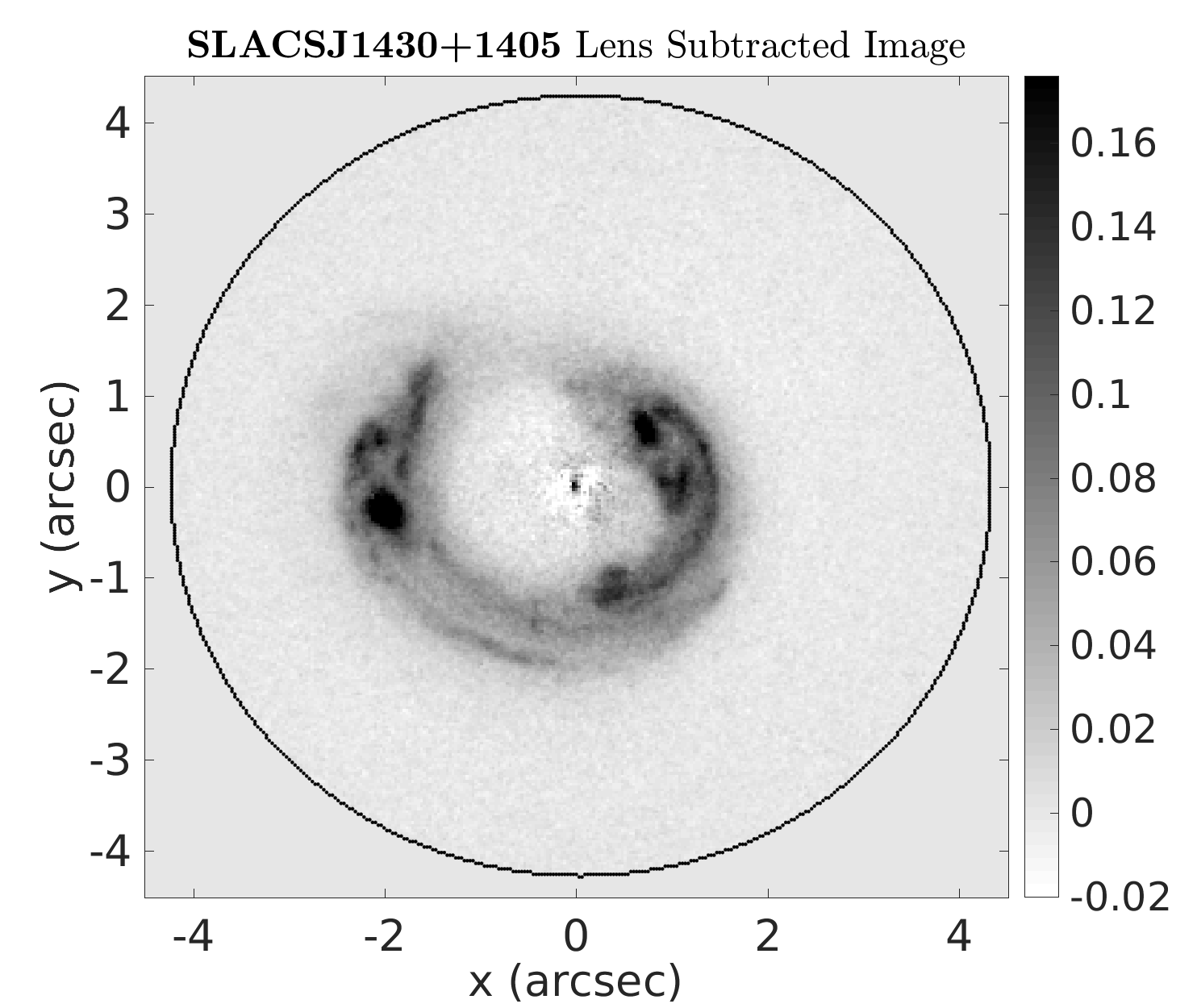}
\includegraphics[width=0.325\textwidth]{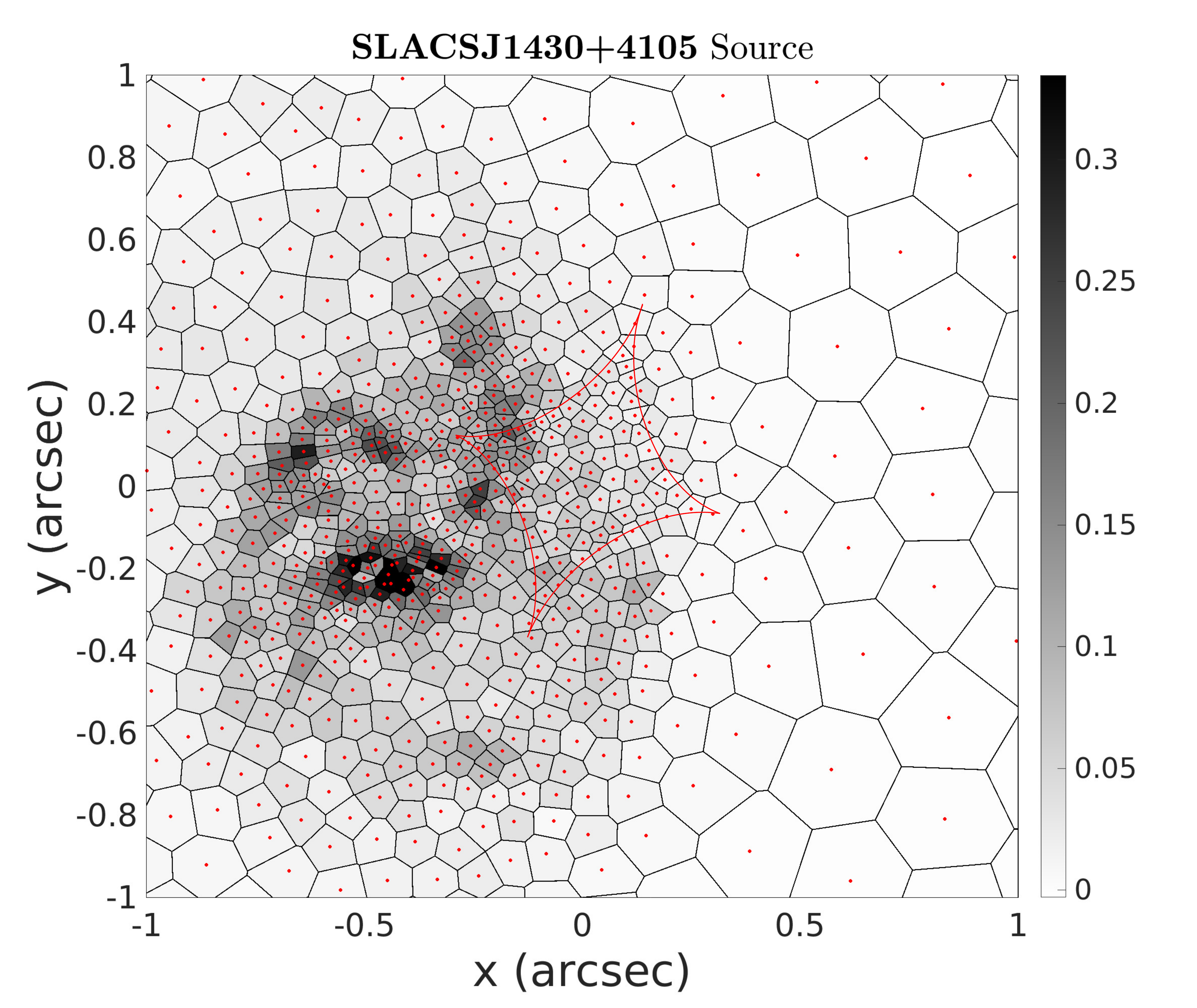}
\caption{The observed images (left column), lens subtracted images (middle column) and source reconstructions (right column) for the three lenses in our sample, SLACSJ0252+0039 (top row), SLACSJ1250+0523 (middle row) and SLACSJ1430+1405 (bottom row). The lens subtractions and source reconstructions are computed using the highest likelihood model found for each image, however their visual appearance does not change significantly for other high likelihood models. The source reconstructions are computed using an adaptive pixellation and are represented as a Voronoi tessellation. The kpc per arcsec scale of each lens is 4.281 kpc/", 3.729 kpc/" and 4.335 kpc/" respectively.} 
\label{figure:Images}
\end{figure*}

The preliminary sample of three lenses used in this work is taken from the Sloan Lens ACS Survey (SLACS, e.g.\ \citealt{Bolton2008}), where galaxies were initially selected from the Sloan Digital Sky Survey (SDSS) main \citep{Strauss2002} and luminous red galaxy \citep[LRG,][]{Eisenstein2001} catalogues, as containing spectral emission lines from more than one redshift in a single fibre. The SLACS lens galaxies therefore primarily consist of ME galaxies with no known differences to other similar galaxies in the main or LRG parent samples \citep[e.g.][]{Bolton2006, Treu2009}, other than a bias towards higher total mass.

Imaging of SLACS lenses is available across a range of wavelengths and spatial resolutions. We analyse monochromatic images, and have two primary considerations in selecting a waveband: (i) high spatial resolution to better constrain the lens model and; (ii) longer wavelength coverage to be more sensitive to the lens galaxy's old stellar populations (which contribute most significantly to its mass). We choose HST/Advanced Camera for Surveys (ACS) imaging taken with the F814W filter, which provides high-resolution imaging at rest-frame near-infrared (NIR) wavelengths. We also require that this imaging is available with an exposure time of at least one Hubble orbit. This work is focused on showing what can be measured for individual systems, so selection effects are not important for this work's conclusions.

These criteria yield a sample of $\sim$40 ME lenses, which SDSS spectroscopy separates into three bins of velocity dispersion (uncorrected for aperture effects): low ($230$$<$$\sigma_\mathrm{SDSS}$$<$$270$~km/s), intermediate ($310$$<$$\sigma_\mathrm{SDSS}$$<$$350$~km/s) and high ($\sigma_\mathrm{SDSS}$$>$$360$~km/s). We select one lens randomly from each bin, resulting in the lenses SDSSJ0252+0039 (low $\sigma_\mathrm{SDSS}$, hereafter SLACS1), SDSSJ1250+0523 (intermediate $\sigma_\mathrm{SDSS}$, hereafter SLACS2) and SDSS1430+4105 (high $\sigma_\mathrm{SDSS}$, hereafter SLACS3). None are cD or Brightest Cluster Galaxies, but SLACS3 is in the field of a nearby cluster \citep{Treu2009}. SLACS1 and SLACS3 were both observed in HST programme GO-10886 and SLACS2 in programme GO-10494. Table \ref{table:SampleProp} gives a summary of each lens's observed properties \citep{Bolton2008}, including $\sigma_\mathrm{SDSS}$, the source and lens redshifts, $z_{src}$ and $z_{lens}$. Table \ref{table:SampleModel} shows the results of previous SLACS analyses -- in particular photometry, stellar population modeling and stellar dynamics \citep{Auger2009, Auger2009a}, and mass modeling \citep{Bolton2008}.

\begin{table*}
\resizebox{\linewidth}{!}{
\begin{tabular}{ l | l | l | l | l | l | l | l | l } 
\multicolumn{1}{p{2.2cm}|}{Target name} 
& \multicolumn{1}{|p{1.5cm}|}{RA} 
& \multicolumn{1}{|p{1.5cm}|}{Dec} 
& \multicolumn{1}{|p{1.6cm}|}{$z_{\rm lens}$} 
& \multicolumn{1}{|p{1.4cm}|}{$z_{\rm src}$} 
& \multicolumn{1}{|p{2.3cm}|}{$\sigma_\mathrm{SDSS}$ (km/s)}  
& \multicolumn{1}{|p{2.3cm}|}{Near a cluster?}  
\\ \hline
& & & & & & & &\\[-4pt]
SDSSJ0252+0039 (SLACS1) & 02$h$\,52$\arcmin$45.21$\arcsec$ & +00$^\circ$39$\arcmin$58$\arcsec$  & 0.2803  & 0.9818  & $164\pm12$ & - \\[2pt]
SDSSJ1250+0523 (SLACS2) & 12$h$\,50$\arcmin$28.26$\arcsec$ & +05$^\circ$23$\arcmin$49$\arcsec$  & 0.2318  & 0.7953  & $252\pm14$ & - \\[2pt]
SDSSJ1430+4105 (SLACS3) & 14$h$\,30$\arcmin$04.10$\arcsec$ & +41$^\circ$05$\arcmin$57$\arcsec$  & 0.2850  & 0.5753  & $322\pm32$ & MaxBCGJ217.49493+41.10435 ($z = 0.270$) \\[2pt]
\end{tabular}
}
\caption{Properties of the three SLACS lenses studied in this paper. Columns 2 and 3 gives their Right Ascension and Declination in J2000 coordinates. Columns 4, 5 and 6 give the redshifts of the lens and source, and the velocity dispersion of the lens (uncorrected for aperture effects) \citep{Bolton2008}. Column 7 notes whether the lens is near a known cluster \citep{Treu2009}.}
\label{table:SampleProp}
\end{table*}

\begin{table*}
\resizebox{\linewidth}{!}{
\begin{tabular}{ l | l | l | l | l | l | l | l | l | l | l | l | l | l | l } 
\multicolumn{1}{p{1.2cm}|}{Target name} 
& \multicolumn{1}{p{1.2cm}|}{$I_{\rm 814}$ (obs.)} 
& \multicolumn{1}{p{1.0cm}|}{$r_{\rm e}$ (")} 
& \multicolumn{1}{p{1.0cm}|}{$q$ (b/a)} 
& \multicolumn{1}{p{1.0cm}|}{$\theta$ ($^{\circ}$)}  
& \multicolumn{1}{p{1.4cm}|}{$log[\frac{M_{Ein}}{M_{\odot}}]$} 
& \multicolumn{1}{p{1.0cm}|}{$b_{\rm SIE}$} 
& \multicolumn{1}{p{1.0cm}|}{$q$ (SIE)} 
& \multicolumn{1}{p{1.0cm}|}{$\theta$ (SIE)} 
& \multicolumn{1}{p{1.0cm}}{$N_{\rm src}$} 
& \multicolumn{1}{p{2.0cm}}{$log[\frac{M_{\rm *}^{Chab}}{M_{\odot}}]$}
& \multicolumn{1}{p{2.0cm}}{$log[\frac{M_{\rm *}^{Sal}}{M_{\odot}}]$} 
& \multicolumn{1}{p{2.0cm}|}{$f_{\rm DM}$ (Chab)} 
& \multicolumn{1}{p{2.0cm}|}{$f_{\rm DM}$ (Sal)} 
& \multicolumn{1}{p{2.0cm}}{$\alpha_{\rm SD}$} 
\\ \hline
& & & & & & & & & & & & & &\\[-4pt]
SDSSJ0252+0039 (SLACS1) & 18.04  & 1.39  & 0.94 & 97.2   & 11.25 & 1.04  & 0.93 & 106.2 & 3 & $11.21\pm0.13$ & $11.46\pm0.13$ & $0.42\pm0.18$ & $-0.03\pm0.32$ & $1.57\pm0.12$ \\[2pt]
SDSSJ1250+0523 (SLACS2) & 16.70  & 1.81  & 0.97 & 114.8  & 11.26 & 1.13  & 0.96 & 130.8 & 5 & $11.53\pm0.07$ & $11.77\pm0.07$ & $0.31\pm0.11$ & $-0.22\pm0.20$ & $2.30\pm0.12$ \\[2pt]
SDSSJ1430+4105 (SLACS3)  & 16.87  & 2.55  & 0.79 & 120.7  & 11.73 & 1.52  & 0.68 & 111.7 & 6 & $11.68\pm0.12$ & $11.93\pm0.11$ & $0.63\pm0.10$ & $ 0.35\pm0.18$ & $2.06\pm0.18$ \\[2pt]
\end{tabular}
}
\caption{Previous measurements of the stellar and mass distributions of the SLACS lenses, taken from the literature. Columns 2-5 give the F814W-band magnitude (non-dust corrected), effective radius, axis ratio and position angle of the lens galaxy, taken from \citet{Bolton2008} and computed by fitting each image with a de Vaucouleurs light profile. Columns 6-9 give the Einstein radius, axis-ratio, position angle and number of source components, taken from \citet{Bolton2008} and inferred by fitting each lens with an SIE mass profile and parametric source. Columns 10-11 give the stellar mass for a Chabrier and Salpeter IMF and are taken from \citet{Auger2009}. Columns 12-13 give the dark matter fractions for a Chabrier IMF and Salpeter IMF from \citet{Auger2009}. Column 14 gives the power-law density slope inferred via a joint lensing and stellar dynamics analysis from \citet{Auger2009}.}
\label{table:SampleModel}
\end{table*}

\subsection{Sample Description}

Based on previous SLACS results (tables \ref{table:SampleProp} and \ref{table:SampleModel}) and the reduced images in figure~\ref{figure:Images} we give a brief description of each lens in our sample. For clarity, figure~\ref{figure:Images} also shows an image of the background galaxies after subtracting the lens light (middle column), and de-lensed reconstructions of that source (right column), which are computed using the highest likelihood lens model found by {\tt PyAutoLens} (which is described next). Neither the lens subtraction nor source reconstruction show large differences when plotted using other high-likelihood lens models, thus these images are indicative of the analysis for any good lens model.

\subsubsection{SLACS1}

SLACS1 is pictured in the top row of figure \ref{figure:Images}. The lens galaxy has an elliptical visual morphology extending smoothly from its centre to  $\sim 3$". The lensed source shows arcs above and below the lens galaxy, which the source reconstruction reveals are a compact knot of light just outside the fold caustic. This lens is therefore a doubly imaged system with minimal extended structure in the source's surface brightness. 

The lens is the lowest mass galaxy in our sample, with an Einstein Mass $M_{\rm Ein} = 11.25 M_{\rm \odot}$ within its Einstein Radius $R_{\rm Ein} = 4.40$kpc, a stellar mass (Chabrier IMF) $M_{\rm \odot}^{\rm Chab} = 11.21\pm0.13 M_{\rm \odot}$ and a velocity dispersion $\sigma_{\rm SDSS} = 164\pm12$ km/s. This translates to a stellar mass fraction within $R_{\rm Ein}$ of $f^{\rm Chab}_{\rm *,Ein} = 0.40\pm0.12$ for a Charbrier IMF. Notably, it has the lowest density slope (inferred via lensing and stellar dynamics) in the whole SLACS sample, with a value $\alpha_{SD} = 1.57 \pm 0.12$.

The HST imaging of SLACS1 had three faint galaxies not associated with the lens or source which overlapped the masked region within which the analysis is performed. These were subtracted using a linear light profile fitting routine and their pixel variances were increased to infinity such that the analysis ignored them. The object at $(x,y) = (3.0", 1.5")$ was included in the lens model as a singular isothermal sphere fixed to these coordinates. However, omitting it was found to have no impact on the results discussed in this work. 

\subsubsection{SLACS2}

SLACS2 is pictured in the middle row of figure \ref{figure:Images}. The lens galaxy again visually shows one smooth extended component, but extending further, to $\sim 3.5$". The lensed source is complex, with a near-circular ring of light, two radial arcs protruding into the lens's centre and six distinct flux peaks. The source reconstruction reveals two galaxies either side of the inner caustic, with features indicative of merging visible in their extended surface brightness profiles, in particular a disrupted tidal tail trailing each source. These tails form the inner ring of light, radial arcs and extended arcs outside the lens in the image-plane. 

Although this lens was chosen from the intermediate $\sigma_{\rm SDSS}$ bin with a value $\sigma_{\rm SDSS} = 252\pm14$ km/s, it turns out to be the highest mass object in our sample with $M_{\rm Ein} = 11.26 M_{\rm \odot}$ within $R_{\rm Ein} = 4.18$kpc and $M_{\rm \odot}^{\rm Chab} = 11.77\pm0.07 M_{\rm \odot}$. Its stellar mass fraction is also the highest in our sample with $f^{\rm Chab}_{\rm *,Ein} = 0.68\pm0.11$. Finally, its density slope $\alpha_{\rm SD} = 2.30 \pm 0.12$ makes it one of the steeper slopes in SLACS and the steepest in our sample.

\subsubsection{SLACS3}

SLACS3 is pictured in the bottom row column of figure \ref{figure:Images}. The lens again shows one smooth extended component. The source shows two bright knots of light in a double image configuration, which the source reconstruction shows is the source galaxy's central bulge. However, also visible is a wealth of additional extended structure surrounding this bulge, which in the image-plane forms multiple giant arcs around the lens. The source may be a face-on spiral galaxy, where the arms on the opposite side of the caustic are not visible due to the reduced magnification. Alternatively, it is a complex merging system.

This object has $M_{\rm Ein} = 11.73 M_{\rm \odot}$ within $R_{\rm Ein} = 6.53$kpc, $M_{\rm \odot}^{\rm Chab} = 11.68\pm0.12 M_{\rm \odot}$ and $\sigma_{\rm SDSS} = 322\pm32$ km/s. It has a stellar mass fraction $f^{\rm Chab}_{\rm *,Ein} = 0.33\pm0.09$ and density slope $\alpha_{\rm SD} = 2.06 \pm 0.18$. Unlike the other lenses in our sample, SLACS3 is near a cluster \citep{Treu2009}.        
\section{Method}\label{Method}

When an extended source is gravitationally lensed, light rays emanating from different regions of the source galaxy trace different paths through the lens galaxy. This provides a projected and extended view of the lens's gravitational potential, information which can be recovered with knowledge of the source's unlensed light distribution. Thus, we are exploiting the lensed source's \textit{surface brightness profile}, in contrast to the previous photometric studies of SLACS lenses which used only the source's \textit{position} to measure the lens galaxy's Einstein Mass, $M_{\rm Ein}$ (e.g.\ \citealt{Auger2009a}). It is this exploitation of the source's extended information which allows us to constrain the lens's underlying mass structure.

To perform this analysis we use our new lens modeling software {\tt PyAutoLens}\footnote{The {\tt PyAutoLens} software is open-source and available from \url{https://github.com/Jammy2211/PyAutoLens}. The results in this work were computed using an earlier {\tt Fortran} build of {\tt AutoLens}}, which is described in \citet[][N18 hereafter]{Nightingale2018}, building on the works of \citet[][WD03 hereafter]{Warren2003}, \citet[][S06 hereafter]{Suyu2006} and \citet[][N15 hereafter]{Nightingale2015}. We refer readers to these works for a full description of {\tt PyAutoLens}.  Key points to note are: \vspace{-2mm}
\begin{enumerate}
    \item the lens galaxy's light and mass distributions are fitted simultaneously;
    \item the source's surface brightness distribution is reconstructed on an adaptive pixel-grid (see the right column of figure \ref{figure:Images});
    \item the Bayesian framework of S06 is used to objectively determine the most probable source reconstruction, and the complexity of the lens model is also chosen objectively via Bayesian model comparison \citep[using {\tt MultiNest};][]{Feroz2009,Feroz2013};
    \item the method is fully automated and requires no user intervention (after a brief initial setup) for the analysis presented in this work.
\end{enumerate}

\begin{table*}
\tiny
\resizebox{\linewidth}{!}{
\begin{tabular}{ l | l | l | l l l l l l l } 
\multicolumn{1}{p{1.8cm}|}{\centering \textbf{Model}} 
& \multicolumn{1}{p{1.1cm}|}{\centering \textbf{Compo \\ nent(s)}} 
& \multicolumn{1}{p{1.1cm}|}{\centering \textbf{Represents}} 
& \multicolumn{1}{p{1.5cm}}{\textbf{Parameters}} 
& \multicolumn{1}{p{1.5cm}}{} 
& \multicolumn{1}{p{1.5cm}}{} 
\\ \hline
& & & &  & & & & & \\[-4pt]
\textbf{Sersic} & Light + & Stellar Matter & ($x_{\rm l}$,$y_{\rm l}$) - profile center & $q_{\rm l}$ - axis ratio & $\theta_{\rm l}$ - orientation \\[2pt]
               &  Mass     & & $I_{\rm l}$ - Intensity & $R_{\rm l}$ - Effective Radius & $n_{\rm l}$ - Sersic index\\[2pt]
               & & & $\Psi_{\rm l}$ - Mass-to-Light Ratio & $\Gamma_{\rm l}$ - Radial Gradient & \\[2pt]
\hline
& & & & & & \\[-4pt]
\textbf{Exponential} ($Exp$) & " & " & \multicolumn{3}{p{4.5cm}}{Identical to Sersic with fixed $n_{\rm l} = 1$}  \\[2pt]
\hline
& & & & & & \\[-4pt]
\textbf{Singular Isothermal} & Mass & Total (Stellar  & ($x$,$y$) - profile center & $q$ - axis ratio & $\theta$ - orientation \\[2pt]
\textbf{Ellipsoid} (SIE)   &      & + Dark Matter) & $\theta_{\rm  E}$ - Einstein Radius \\[2pt]
\hline
& & & & & & \\[-4pt]
\textbf{Singular Isothermal Sphere} (SIS) & " & " & \multicolumn{3}{p{6.0cm}}{Identical to SIE with fixed $q = 1$ and $\theta$ omitted} \\[2pt]
\hline
& & & & & & \\[-4pt]
\textbf{Spherical NFW} ($NFWSph$) & " & Dark Matter & ($x_{\rm d}$,$y_{\rm d}$) - profile center & & \\[2pt]
            &   &   & $\kappa_{\rm d}$ - Halo normalization & \multicolumn{2}{p{3.5cm}}{$r_{\rm s}$ - Scale radius (fixed to 30 kpc)} \\[2pt]
\hline
& & & & & & \\[-4pt]
\textbf{Elliptical NFW} ($NFWEll$) & " & " & ($x_{\rm d}$,$y_{\rm d}$) - profile center & $q_{\rm d} - profile axis-ratio $ & \\[2pt]
            &   &   & $\kappa_{\rm d}$ - Halo normalization & \multicolumn{2}{p{3.5cm}}{$r_{\rm s}$ - Scale radius (fixed to 30 kpc)} \\[2pt]
\hline
& & & & & & \\[-4pt]
\textbf{Generalized spherical NFW} ($gNFWSph$) & " & " & ($x_{\rm d}$,$y_{\rm d}$) - profile center & $\gamma_{\rm d}$ - inner slope & \\[2pt]
            &   &   & $\kappa_{\rm d}$ - Halo normalization & \multicolumn{2}{p{3.5cm}}{$r_{\rm s}$ - Scale radius (fixed to 30 kpc)} \\[2pt]
\hline
& & & & & & \\[-4pt]
\textbf{Shear} & " & Line-of-sight & $\gamma_{\rm sh}$ - magnitude & $\theta_{\rm sh}$ - orientation & \\[2pt] 
\hline
& & & & & & \\[-4pt]
\end{tabular}
}
\caption{The light and mass profiles used to model each strong lens. Column 1 gives the model name. Column 2 whether it models the lens's light and mass or just its mass. Column 3 states which mass component it represents. Column 4 gives its associated parameters, where all orientation angles $\theta$ are defined counter-clockwise from the positive x-axis.}
\label{table:SimModels}
\end{table*}

Table~\ref{table:SimModels} summarizes the lens models that are available to fit the lenses in our sample, alongside the free parameters associated with each. The equations and a description of each profile are given in appendix \ref{LensModels}. In brief, a light profile is used to generate an intensity map $I$, which fits and subtracts the lens's light. A mass profile is used to generate a surface density profile, $\kappa (x)$, which is integrated to compute a deflection angle map $\vec{{\alpha}}_{\rm x,y}$, allowing one to ray-trace image-pixels to the source-plane for the source reconstruction. When multiple light or mass profiles are modeled simultaneously, their individual intensities or deflection angles are summed and their parameters are given numerical subscripts (e.g.\ $x_{\rm l1}$, $n_{\rm l2}$, etc.).

{\tt PyAutoLens} fits the lens galaxy's light whilst simultaneously modeling its mass and reconstructing the source. Therefore, before each source reconstruction, a light profile for the lens is computed, convolved with the instrumental PSF and subtracted from the observed image, with the source analysis subsequently performed on this residual image. The light and mass models are therefore sampled from the same non-linear parameter space, meaning that the Sersic profile(s) used to fit the lens's light can be translated to stellar density profiles (assuming a mass-to-light profile) and incorporated into the mass model. Thus, the light profiles may be constrained by both their fit to the lens galaxy's light and the strongly lensed image of the source galaxy. We can therefore take two independent approaches to lens modeling which make different assumptions about the lens's total mass distribution:

\begin{itemize}

\item \textit{Total Mass Model:} the model represents all of the lens galaxy's mass (i.e.\ both stellar and dark matter), and the lens galaxy's light profile is not used to constrain the mass model. Thus, the lens's light profile is constrained only by a fit to the surface photometry of the galaxy (and not by the strong lensing analysis), as in the previous studies of galaxy structure referenced in the introduction. The total mass model used in this work is a singular isothermal ellipsoid ($SIE$).

\end{itemize}

\begin{itemize}

\item \textit{Decomposed Mass Model}: the model splits the lens galaxy's mass into stellar and dark matter components, with the lens galaxy's light profile included in the former. Thus, the lens's light profile is constrained by both its fit to the lens galaxy's light and by the strong lensing analysis -- thereby requiring that each component of the light profile corresponds to a genuine mass structure. The decomposed mass models used in this work assume either a single elliptical $Sersic$ or two-component $Sersic$ (with Sersic index free)  + Exponential ($Exp$, with Sersic index fixed to $1$) profile for the stellar mass. A spherical Navarro-Frenk-White ($NFWSph$) profile is used for the dark matter. Our choice of an $Exp$ in the two component model is not an attempt to represent a specific galaxy structure (e.g. a disk). Instead, it is motivated by model comparison, which will be shown in the next section.

\end{itemize}

The results section of this work will compare both of the approaches above, to quantify the impact of including strong lensing information in measurements of galaxy structure. Both of the mass models above include an external shear term, which accounts for the contribution of line-of-sight structures to the strong lensing signal.

We stress that we are \textit{not} comparing the total-mass model and decomposed mass models so as to try and determine which `better' fits the lensing data, nor do we compare their inferred density profiles in this work. Such a comparison would require us to fit a more complex total-mass profile (e.g. with a variable inner density slope) and to consider well-known degeneracies like the source-position transformation \citep{Schneider2013}. Our use of a total-mass model is motivated only to fit the lens galaxy's light in a way that is independent of the mass model, as shown in N18.

When multiple light profiles are fitted (e.g.\ $Sersic$ + $Exp$), one has to make a decision as to whether certain parameters that are shared by both components are independent of one another, like their centers ($x_{\rm l}$ and $y_{\rm l}$), orientations ($\theta_{\rm l}$) and mass-to-light ratios ($\Psi_{\rm l}$). Rather than making ad hoc assumptions about these parameters, {\tt PyAutoLens} uses Bayesian model comparison to objectively determine the lens model complexity. At different phases of the analysis, different parameterizations of the light and mass models are fitted using multiple independent non-linear searches, computing the Bayesian evidence of each model via {\tt MultiNest}. We favour a more complex model when its Bayes factor (ratio of Bayesian evidences) is twenty or above (considered `strong' evidence in Bayesian statistics). Thus, the choice of lens model incorporates the principle of Occam's razor, where a more complex model not only has to fit the data better, but better enough to justify its extra parameters and complexity. N18 demonstrated that this is an effective way to choose the appropriate model complexity, in simulations where the correct answer is known. The specifics of each model comparison phase are described next in their corresponding results section.

When quoting the best-fit lens models, the `most-probable' set of parameters are quoted, corresponding to the median of their one-dimensional marginalized posterior probability distribution. Errors are quoted at $3\sigma$, corresponding to the $0.3$ and $99.7$ percentiles of this posterior.   
\section{Results}\label{Results}

We now describe how Bayesian model comparison informs the model of each lens galaxy. We quote the increase in evidence between two models, $\Delta$ln$\epsilon = $ln$\epsilon_{\rm model1} - $ln$\epsilon_{\rm model2}$, referred to as the Bayes factor. In Bayesian statistics, a Bayes factor of $10$ would roughly correspond to a `substantial' preference of model $1$, whereas our criterion of $20$ is considered `decisive'.

The overall normalization of the evidence depends on how {\tt PyAutoLens}'s adaptive analysis is initialized (see N18), which is performed independently for the total-mass and decomposed-mass profile analyses. This means the evidence values ln$\epsilon$ for these two profiles \textit{cannot be compared directly} (which as discussed in section \ref{Method}, is not something we perform in this work anyway). Instead, a fair comparison must compare $\Delta$ln$\epsilon$ between different models of the total-mass profile and decomposed-mass profiles. For example, in the next section, we compare $\Delta$ln$\epsilon$ for the $Sersic$ and $Sersic$ + $Exp$ models for both the total-mass and decomposed profiles, but we do not compare $\Delta$ln$\epsilon$ values between the total-mass profile and decomposed mass profiles themselves. The inferred parameters for individual lens models are discussed throughout this section and the tables containing their values are at the end of the script in appendix \ref{LensResults}. 

\subsection{Light Profile Components}\label{LightProfMC}

The number of light profile components is chosen by comparing two models, a single $Sersic$ profile and a two-component $Sersic$ + $Exp$ model where the two components are centrally and rotationally aligned ($x_{\rm l1} = x_{\rm l2}$, $y_{\rm l1} = y_{\rm l2}$ and $\theta_{\rm l1} = \theta_{\rm l2}$). We perform two independent runs of this phase, for each image, using: (i) a total mass profile ($SIE$); (ii) a decomposed mass profile ($Light$ + $NFWSph$) with shared mass-to-light ratio ($\Psi_{\rm l1} = \Psi_{\rm l2}$) between the multiple light components. The results are presented in table \ref{table:LightMC} and address the following question:

\begin{itemize}

\item \textbf{Are the lenses single component or multi-component systems?}

For all three images, the single $Sersic$ model gives a significantly lower Bayesian evidence than the multi-component model. This occurs regardless of whether we use the total mass or decomposed mass model. For the total-mass profile, the Bayes factors preferring the two-component models are $\Delta$ln$\epsilon = 568$ for SLACS1, $\Delta$ln$\epsilon = 224$ for SLACS2 and $\Delta$ln$\epsilon = 297$ for SLACS3. For the decomposed models they are $\Delta$ln$\epsilon = 539$ for SLACS1, $\Delta$ln$\epsilon = 106$ for SLACS2 and $\Delta$ln$\epsilon = 535$ for SLACS3. Thus, the inclusion of lensing does not change the number of physical components we associate with each galaxy. However, it leads to lower Bayes factors in SLACS1 ($-29$) and SLACS2 ($-118$) and a higher factor in SLACS3 ($238$).

\end{itemize}

The parameters of the light profiles fitted in this section are given in tables \ref{table:SingleSersicParams} and \ref{table:TwoCompAlignParams}. For each lens, the single $Sersic$ model goes to values of $n_{\rm l1} > 3.5$ and non-circular axis-ratios ($q_{\rm l1}$ around 0.9), both typical of massive elliptical galaxies. For the two component model, the first component is centrally compact (lower $R_{\rm l1}$), concentrated (higher $n_{\rm l1}$) and round ($q_{\rm l1}$ near 1.0), whereas the second component is more extended and flatter. 

By comparing the parameters of the total-mass and decomposed-mass models given in tables \ref{table:SingleSersicParams} and \ref{table:TwoCompAlignParams}, we can determine whether the use of lensing changes the lens galaxy's inferred light profile and therefore structure. For both the $Sersic$ and $Sersic$ + $Exp$ models, consistent fits are obtained either by a photometric-only analysis, or one that includes lensing. Thus, our use of lensing information has (in the analysis so far) not affected the light profile we infer (as might be expected given $\Delta$ln$\epsilon$ in SLACS1 and SLACS2 was lower for the decomposed-mass profile than the total-mass profile). There are two ways one can interpret this. On the one hand, it could indicate that the light profile fits from photometry alone are a reliable tracer of the lens's underlying mass structure. On the other hand, if the assumed light profile is not a representative description of the lens's underlying mass structure, lensing would be unable to change the model in a way that alters the inferred light profile anyway.

In table \ref{table:LightMCSersic} we compare a $Sersic$ + $Exp$ and $Sersic$ + $Sersic$ model, using both the total-mass and decomposed-mass profiles. For all images, we find that the $Sersic$ + $Exp$ model provides a higher Bayesian evidence than the $Sersic$ + $Sersic$ model. Furthermore, the Sersic index's of the second Sersic components, $n_{\rm l2}$, are all consistent with $n_{\rm l2} = 1.0$. Therefore, the $Sersic$ + $Exp$ model is used hereafter.

\begin{table*}
\resizebox{\linewidth}{!}{
\begin{tabular}{ l | l | l | l | l | l } 
\multicolumn{1}{p{3.0cm}|}{\centering Target Name} 
& \multicolumn{1}{p{1.5cm}|}{\centering Mass \\ Profile} 
& \multicolumn{1}{p{1.6cm}|}{Mass-to-light Ratio} 
& \multicolumn{1}{p{2.6cm}|}{Second Light Component Profile} 
& \multicolumn{1}{p{1.8cm}|}{Single Sersic}  
& \multicolumn{1}{p{2.0cm}|}{\centering Sersic + Halo \\ (shared geometry)}  
\\ \hline
& & & & & \\[-4pt]
$\mathbf{SLACSJ0252+0039}$ & Total & N/A & Exponential & 212290.8 & \textbf{212858.4} \\[2pt]
$\mathbf{SLACSJ0252+0039}$ & Decomposed & Shared & Exponential & 212273.4 & \textbf{212812.6} \\[2pt]
\hline

$\mathbf{SLACSJ1250+0523}$ & Total & N/A & Exponential & 216990.6 & \textbf{217214.5} \\[2pt]
$\mathbf{SLACSJ1250+0523}$ & Decomposed & Shared & Exponential & 217074.0 & \textbf{217180.3}  \\[2pt]
\hline

$\mathbf{SLACSJ1430+4105}$ & Total & N/A & Exponential & 217423.3 & \textbf{217720.7} \\[2pt]
$\mathbf{SLACSJ1430+4105}$ & Decomposed & Shared & Exponential & 217110.8 & \textbf{217645.2} \\[2pt]
\end{tabular}
}
\caption{The Bayesian Evidence of each model used in the light profile model comparison phase, comparing the $Sersic$ and $Sersic$ + $Exp$ models for the lens galaxy light profile. For each image, this phase computes two models, assuming: (i) a total mass $SIE$ profile (rows 1, 3 and 5); (ii) a decomposed $Light$ + $NFWSph$ profile (rows 2, 4 and 6). Column 1 gives each image's target name, column 2 the type of mass profile, column 3 the mass-to-light ratio assumption, column 4 the profile of the second light component (all Exponential in this table) and columns 5-6 the Bayesian Evidence values of each model, where bolded values show those chosen via model comparison}
\label{table:LightMC}
\end{table*}

\begin{table*}
\resizebox{\linewidth}{!}{
\begin{tabular}{ l | l | l | l | l | l } 
\multicolumn{1}{p{3.0cm}|}{\centering Target Name} 
& \multicolumn{1}{p{1.5cm}|}{\centering Mass \\ Profile} 
& \multicolumn{1}{p{1.6cm}|}{Mass-to-light Ratio} 
& \multicolumn{1}{p{2.6cm}|}{Second Component Exponential} 
& \multicolumn{1}{p{1.8cm}|}{Second Component Sersic}  
& \multicolumn{1}{p{2.0cm}|}{Sersic Index}  
\\ \hline
& & & & & \\[-4pt]
$\mathbf{SLACSJ0252+0039}$ & Total & N/A & \textbf{212858.4} & 212855.5 & $n_{\rm l2} = 1.16^{+0.23}_{-0.32}$ \\[2pt]
$\mathbf{SLACSJ0252+0039}$ & Decomposed & Shared & \textbf{212812.6} & 212791.4 & $n_{\rm l2} = 1.15^{+0.27}_{-0.29}$ \\[2pt]
\hline

$\mathbf{SLACSJ1250+0523}$ & Total & N/A & \textbf{217214.5} & 217189.2 & $n_{\rm l2} = 1.23^{+0.24}_{-0.26}$ \\[2pt]
$\mathbf{SLACSJ1250+0523}$ & Decomposed & Shared & \textbf{217180.3} & 217172.2 & $n_{\rm l2} = 1.26^{+0.24}_{-0.29}$ \\[2pt]
\hline

$\mathbf{SLACSJ1430+4105}$ & Total & N/A & \textbf{217720.7} & 217721.3 & $n_{\rm l2} = 0.86^{+0.27}_{-0.30}$ \\[2pt]
$\mathbf{SLACSJ1430+4105}$ & Decomposed & Shared & \textbf{217645.2} & 217640.2 & $n_{\rm l2} = 1.03^{+0.24}_{-0.36}$ \\[2pt]
\end{tabular}
}
\caption{The Bayesian Evidence of each model used in the light profile model comparison phase, comparing the $Sersic$ + $Exp$ and $Sersic$ + $Sersic$ models for the lens galaxy light profile. For each image, this phase computes two models, assuming: (i) a total mass $SIE$ profile (rows 1, 3 and 5); (ii) a decomposed $Light$ + $NFWSph$ profile (rows 2, 4 and 6). Column 1 gives each image's target name, column 2 the type of mass profile, column 3 the mass-to-light ratio assumption and columns 4-5 the Bayesian Evidence values of each model, where bolded values show those chosen via model comparison. The 6th column shows the Sersic index inferred for the second component of the $Sersic$ + $Sersic$ model.}
\label{table:LightMCSersic}
\end{table*}

\subsection{Light Profile Geometry}\label{LightProfMC}

\begin{table*}
\resizebox{\linewidth}{!}{
\begin{tabular}{ l | l | l | l | l | l | l | l } 
\multicolumn{1}{p{1.2cm}|}{Target Name} 
& \multicolumn{1}{p{1.5cm}|}{\centering Mass \\ Profile} 
& \multicolumn{1}{p{1.6cm}|}{Mass-to-light Ratio} 
& \multicolumn{1}{p{1.5cm}|}{Second Light Component Profile} 
& \multicolumn{1}{p{1.5cm}|}{\centering Sersic + Halo \\ (shared geometry)}  
& \multicolumn{1}{p{1.5cm}|}{\centering Sersic + Halo \\ (independent $\theta_{\rm l}$)}  
& \multicolumn{1}{p{1.5cm}|}{\centering Sersic + Halo \\ (independent $x_{\rm l}$ / $y_{\rm l}$)}  
& \multicolumn{1}{p{1.5cm}|}{\centering Sersic + Halo \\ (independent geometry)}  
\\ \hline
& & & & & & &\\[-4pt]
$\mathbf{SLACSJ0252+0039}$ & Total & N/A & Exponential & 212858.4 & \textbf{212912.7} & 212870.2 & 212890.7 \\[2pt]
$\mathbf{SLACSJ0252+0039}$ & Decomposed & Shared & Exponential & 212812.6 & \textbf{212865.5} & 212804.1 & 212871.3 \\[2pt]
$\mathbf{SLACSJ0252+0039}$ & Decomposed & Independent & Exponential & 212819.4 & \textbf{212867.3} & 212861.7 & 212862.5 \\[2pt]
\hline

$\mathbf{SLACSJ1250+0523}$ & Total & N/A & Exponential & 217214.5 & 217228.1 & 217223.5 & \textbf{217249.0} \\[2pt]
$\mathbf{SLACSJ1250+0523}$ & Decomposed & Shared & Exponential & 217180.3 & 217194.0 & 217210.9 & \textbf{217219.7} \\[2pt]

$\mathbf{SLACSJ1250+0523}$ & Decomposed & Independent & Exponential & 217183.2 & 217216.6 & 217205.2 & \textbf{217240.4} \\[2pt]
\hline

$\mathbf{SLACSJ1430+4105}$ & Total & N/A & Exponential & 217720.7 & \textbf{217840.3} & 217753.2 & 217852.4 \\[2pt]
$\mathbf{SLACSJ1430+4105}$ & Decomposed & Shared & Exponential & 217645.2 & 217780.8 & 217663.0 & \textbf{217804.7} \\[2pt]
$\mathbf{SLACSJ1430+4105}$ & Decomposed & Independent & Exponential & 217676.5 & 217823.4 & 217694.5 & \textbf{217876.0} \\[2pt]
\end{tabular}
}
\caption{The Bayesian Evidence of each model used in the light profile geometry model comparison phase, comparing different light profile geometries for the lens galaxy light profile. For each image, the phase is run three times using a model which: (i) assumes a total-mass $SIE$ profile (rows 1, 4 and 7); (ii) assumes a decomposed $Light$ + $NFWSph$ mass profile where light components share their mass-to-light ratio (rows 2, 5 and 8) and; (iii) where these mass-to-light ratios are instead independent (rows 3, 6 and 9). Column 1 gives each image's target name, column 2 the type of mass profile, column 3 the mass-to-light ratio assumption, column 4 the profile of the second light component (all Exponential in this table) and columns 5-8 the Bayesian Evidence values of each light profile geometry, where bold-face values show those chosen via model comparison}
\label{table:TwoCompOffset}
\end{table*}

Given a model with two components of stellar material is preferred for all three lenses, we next investigate whether these two components are geometrically aligned or offset. This is performed by comparing four $Sersic$ + $Exp$ models where their geometry is: (i) aligned (the results of the previous phase); (ii) offset rotationally ($\theta_{\rm l1} \neq \theta_{\rm l2}$); (iii) offset centrally ($x_{\rm l1} \neq x_{\rm l2}$ and $y_{\rm l1} \neq y_{\rm l2}$) and; (iv) offset in both. We perform three independent runs of this phase, using a total-mass profile ($SIE$) or decomposed mass profile ($Light$ + $NFWSph$) with mass-to-light ratio(s) between the two light components shared ($\Psi_{\rm l1} = \Psi_{\rm l2}$) or independent ($\Psi_{\rm l1} \neq \Psi_{\rm l2}$). These results are presented in table \ref{table:TwoCompOffset} and address the following two questions:

\begin{itemize}

\item \textbf{Are the two components rotationally offset from one another?}

For each lens in table \ref{table:TwoCompOffset} there are two models that include a rotational offset: one where the centers are aligned (column 6) and one where they are offset (column 8). We quote the largest Bayes factors from either column. For all three lenses, rotational offsets are detected for both the total mass model ($\Delta$ ln$\epsilon = 54$, $\Delta$ ln$\epsilon = 35$ and $\Delta$ ln$\epsilon = 120$) and decomposed model ($\Delta$ ln$\epsilon = 53$, $\Delta$ ln$\epsilon = 57$ and $\Delta$ ln$\epsilon = 200$). The inclusion of lensing therefore increases the significance of two detections. 

\item \textbf{Are the two components centrally offset from one another?}

As for the rotational offsets, we quote here the largest Bayes factors available from columns 7 and 8 of table \ref{table:TwoCompOffset}. For the total-mass models, we detect a centroid offset in SLACS2, with a Bayes factor $\Delta$ ln$\epsilon = 21$. This increases to $\Delta$ ln$\epsilon = 24$ for the decomposed model with independent mass-to-light ratios, which also gives a detection in SLACS3 with Bayes factor $\Delta$ ln$\epsilon = 53$. This is the first case where the decomposed-mass model has deviated from the total-mass model.

\end{itemize}

\begin{figure*}
\centering
\includegraphics[width=0.95\textwidth]{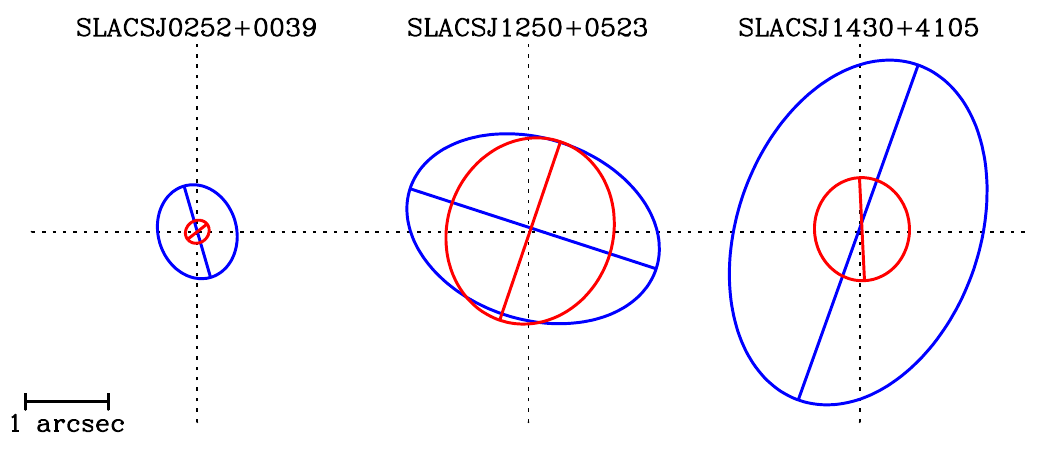}
\caption{A schematic of the mass distribution in each lens galaxy, using the geometrically offset $Sersic$ + $Exp$ model chosen for each lens in table \ref{table:TwoCompOffset}. The red represents the $Sersic$ component and blue the $Exp$ component and each ellipse is drawn at the half-light radius $R_{\rm l}$ of each component. These figures use the parameters of the decomposed-mass profile, but as shown in table \ref{table:TwoCompMisalignParams} the total-mass profile gives consistent values.} 
\label{figure:Geometry}
\end{figure*}

The inferred model parameters are given in table \ref{table:TwoCompMisalignParams}. The detected rotational offsets are of order $\Delta \theta \approx 70^\circ$ for SLACS1, $\Delta \theta \approx 80^\circ$ for SLACS2 and $\Delta \theta \approx 23^\circ$ for SLACS3. SLACS2 and SLACS3 showed centroid offsets of order $\Delta r \approx 0.09$ kpc and $\Delta r \approx 0.344$ kpc respectively. Table \ref{table:TwoCompMisalignParams} confirms that the same geometric offsets are inferred between the two components, regardless of whether a total or decomposed mass model is assumed (even for SLACS3, where the centroid offset's Bayes factor was below $20$ for the total-mass model). Thus, the inclusion of lensing is not changing the model that we infer, but it is reaffirming that what we see in lens galaxy's light is physically genuine.

Figure \ref{figure:Geometry} shows the projected geometry of the $Sersic$ + $Exp$ model in each lens. In SLACS1 and SLACS3 the $Exp$ component can be seen to extend to much larger radii than the $Sersic$ component, whereas in SLACS2 the inner component is of comparable size. 

We now compare the geometry of the total-mass model's $SIE$ profile to that of the lens's two-component light profiles, using the parameters given in table \ref{table:TwoCompMisalignParams}. For SLACS1 and SLACS3, the $SIE$ fit assumes an orientation close to that of the extended $Exp$ component. Next, we will show that the $Exp$ component dominates their inner mass distribution, thus the $SIE$ profile is simply tracing the lens's overall mass. For SLACS2 (where the $Exp$ component contributes less to $M_{\rm Ein}$), the $SIE$ model assumes an orientation $\theta = 106^{\circ}$, halfway between the two light profile components ($\theta_{l1} = 71^{\circ}$ and $\theta_{l2} = 146^{\circ}$). It also appears more spherical than these two components ($q = 0.96$, compared to $q_{l1} = 0.88$ and $q_{l2} = 0.87$), suggesting the best-fit $SIE$ geometry is a compromise between the two components of the decomposed-mass model. The large centroid offset found in SLACS3 shows the same behaviour, whereby the total-mass profile's centre ($x = 0.013"$ and $y = 0.018"$) is half-way between that of the two-component light model ($x_{l1} = 0.03"$, $y_{l1} = 0.03"$, $x_{l2} = -0.01"$ and $y_{l2} = -0.03"$).

\subsection{Mass-To-Light Profile}

\begin{table*}
\resizebox{\linewidth}{!}{
\begin{tabular}{ l | l | l | l | l | l | l | l } 
\multicolumn{1}{p{1.2cm}|}{Target Name} 
& \multicolumn{1}{p{1.5cm}|}{\centering Mass \\ Profile} 
& \multicolumn{1}{p{1.5cm}|}{Independent $\Psi_{\rm l}$?}  
& \multicolumn{1}{p{1.6cm}|}{Rotationally Offset?} 
& \multicolumn{1}{p{1.5cm}|}{Centrally Offset?} 
& \multicolumn{1}{p{1.5cm}|}{\centering No Radial Variation}
& \multicolumn{1}{p{1.5cm}|}{\centering Radial Variation \\ $\Gamma_{\rm l}$ ($Sersic$)}  
& \multicolumn{1}{p{1.5cm}|}{\centering Radial Variation \\ $\Gamma_{\rm l}$ ($Exp$)}  
\\ \hline
& & & & & & &\\[-4pt]
$\mathbf{SLACSJ0252+0039}$ & Decomposed & Yes & Yes & No & \textbf{212867.3} & 212867.4 & 212847.4 \\[2pt]
\hline

$\mathbf{SLACSJ1250+0523}$ & Decomposed & Yes & Yes & Yes & 217246.6 & 217276.0 & \textbf{217300.9} \\[2pt]
\hline

$\mathbf{SLACSJ1430+4105}$ & Decomposed & Yes & Yes & Yes & \textbf{217922.6} & 217918.8 & 217931.1 \\[2pt]
\end{tabular}
}
\caption{The Bayesian Evidence of each model used to test for a radial variation in the mass-to-light profile of each lens. For each image, the phase is run three times using a model which: (i) assumes the same light profile geometry and mass-to-light ratios as determined in the previous phase (column 6); (ii) assume this model again, but with a radial variation in the mass-to-light profile $\Gamma_{\rm}$ of the $Sersic$ component (column 7) and; (iii) in the $Exp$ component (column 8). Column 1 gives each image's target name, column 2 the type of mass profile, column 3 the mass-to-light ratio assumption, column 4 if their is a rotational offset between the $Sersic$ and $Exp$ components and column 5 if there is a centroid offset. Columns 6-8 give the Bayesian Evidence of each model, where bolded values show the model chosen.}
\label{table:RadGrad}
\end{table*}

Next, we establish whether the two components of each lens share the same mass-to-light ratio and whether this ratio varies with radius. The results of using one or two mass-to-lights ratios was presented in section \ref{LightProfMC} and is shown in table \ref{table:TwoCompOffset}. This addresses the following question:

\begin{itemize}

\item \textbf{Do the two components have different mass-to-light ratios?}

By comparing rows 2 and 3, 5 and 6, and 8 and 9 of table \ref{table:TwoCompOffset} we can infer whether the mass-to-light ratios of the two components are the same. For SLACS1, we find no evidence for different mass-to-light ratios, whereas for SLACS2 and SLACS3 we prefer the model with independent $\Psi_{\rm l}$'s with Bayes factors of $\Delta$ ln$\epsilon = 21$ and $\Delta$ ln$\epsilon = 72$.

\end{itemize}

Next, we run a new model comparison phase to investigate the radial variation, comparing two models (which assume the geometric offsets and mass-to-light ratio parametrizations preferred in the previous phases): (i) a model introducing a radial gradient $\Gamma_{\rm l}$ in the mass-to-light ratio of the $Sersic$ component; (ii) a model introducing a radial gradient in the $Exp$ component. The results for this phase (alongside the no radial gradient cases) are given in table \ref{table:RadGrad}, addressing the following question:

\begin{itemize}

\item \textbf{Does the mass in either component vary radially compared to its light?}

Table \ref{table:RadGrad} shows that support for a radial gradient was not found in SLACS1 or SLACS3, but was in SLACS2 in the $Sersic$ component with an evidence of $\Delta$ ln$\epsilon = 30$ and the $Exp$ component with an evidence $\Delta$ ln$\epsilon = 54$. These detections signify a decrease of mass relative to light in the lens's central regions, leading to a shallower inner mass profile. We therefore assume they are indicating the same effect and choose the highest evidence detection which is in the $Exp$ component. 

\end{itemize}

The mass-to-light ratio of the first component $\Psi_{\mathrm{l1}}$ and second component $\Psi_{\mathrm{l2}}$ of each decomposed mass model are given in table \ref{table:MassParams}. For SLACS1, the inferred values of  $\Psi_{\mathrm{l1}}$ and $\Psi_{\mathrm{l2}}$ are consistent with one another, whereas for SLACS3 they are not, as expected from the results of model comparison. In SLACS3, when only one shared value of $\Psi_{\mathrm{l}}$ is assumed, $\Psi_{\mathrm{l}}$ is forced to a value that is not consistent with the inferred value of $\Psi_{\mathrm{l2}}$. This negatively impacts the inferred mass model and highlights that lensing can distinguish the mass-to-light ratios of individual galaxy components.

For SLACS2, independent mass-to-light ratios were preferred by model comparison, but table \ref{table:MassParams} shows they are consistent with one another. This is because, when this model was preferred, the radial gradient $\Gamma_{\rm l}$ was not included in the mass model. Thus, the omission of the radial gradient leads us to falsely prefer a model in which $\Psi_{\mathrm{l1}}$ and $\Psi_{\mathrm{l2}}$ are different. The negative value of $\Gamma_{\rm l2}$ indicates the mass model is placing less mass in the centre of SLACS2. This may explain why, for SLACS2, the increases in evidence throughout the previous phases were typically much lower than SLACS1 or SLACS3, or resulted in decreases of $\Delta$ln$\epsilon$. Decomposed mass models assuming a constant mass-to-light profile were unable to obtain a sufficiently shallow density profile in the galaxy's central regions, resulting in a worse fit compared to the total-mass profile. SLACS2 is the only lens whose source galaxy light extends all the way to its very centre ($< 1$ kpc) in the form of two radial arcs. This may be why we are sensitive to its inner mass profile and thus able to detect a radial gradient, but find no constraints for SLACS1 or SLACS3.

\subsection{Dark Matter Profile}\label{DarkMatter}

\begin{table*}
\resizebox{\linewidth}{!}{
\begin{tabular}{ l | l | l | l | l | l} 
\multicolumn{1}{p{1.2cm}|}{Target Name} 
& \multicolumn{1}{p{1.8cm}|}{\centering Mass \\ Profile}
& \multicolumn{1}{p{1.8cm}|}{$NFWSph$}
& \multicolumn{1}{p{1.8cm}|}{$NFWEll$ (free $\theta_{\rm d}$)}  
& \multicolumn{1}{p{1.8cm}|}{$NFWEll$ (free $x_{\rm d}$ \& $y_{\rm d}$)}  
& \multicolumn{1}{p{1.8cm}|}{$gNFWSph$} 
\\ \hline
& & & & &\\[-4pt]
$\mathbf{SLACSJ0252+0039}$ & Decomposed & \textbf{212854.8} & 212860.1 & 212853.7 & 212852.8  \\[2pt]
\hline

$\mathbf{SLACSJ1250+0523}$ & Decomposed & \textbf{217246.7} & 217259.5 & 217264.2  & 217255.9 \\[2pt]
\hline

$\mathbf{SLACSJ1430+4105}$ & Decomposed & \textbf{217922.0} & 217925.5  & 217928.6 & 217930.1 \\[2pt]
\end{tabular}
}
\caption{The Bayesian Evidence of models testing the use of alternative dark matter $NFW$ profiles, for each lens. All phases assume the same light profile geometry and mass-to-light ratios as determined in the previous phase (except the mass-to-light gradient is omitted for SLACS2). For each image, the phase is run four times assuming a: (i) spherical $NFWSph$ profile; (ii) an elliptical $NFWEll$ profile with rotational angle free to vary and centre fixed to the $Sersic$ component; (iii) a spherical $NFWSph$ with centre free to vary and (iv) a spherical generalized NFW $gNFWSph$ profile. Column 1 gives each image's target name and column 2 the type of mass profile. Columns 3-6 give the Bayesian Evidence of each model, where bolded values show the model chosen.}
\label{table:DarkMatter}
\end{table*}

Next, we run a model comparison phase assuming more complex models for the dark matter halo of each lens galaxy. This will tell us whether we can go beyond the spherical $NFWSph$ model and inform us what impact the assumption of a relatively simple dark matter model has on our inferred stellar mass density profile. Both model comparisons use the mass-to-light profile chosen in the previous section, however the mass-to-light gradient chosen for $SLACS2$ is omitted as the complexity of this parameter space was too difficult to sample robustly. The model comparison is composed of four phases: (i) a $NFWSph$ (the result of the previous phase); (ii) an elliptical $NFW$ profile, $NFWEll$, whose centred is fixed to the $Sersic$ component but whose rotational angle is free to vary; (iii) a $NFWSph$ where the centre is free to vary (and rotational angle fixed to the $Sersic$ componenets) and; (iv) a spherical generalized $NFW$ profile, $gNFWSph$. These results are presented in table \ref{table:DarkMatter} and address the following two questions:

\begin{itemize}

\item \textbf{What do we learn about the dark matter geometry?}

Table \ref{table:DarkMatter} shows that none of the models using an elliptical $NFW$ profile produce a Bayes factor above $20$, meaning that there was insufficient information in the lens data to constrain the ellipticity of the dark matter halo or detect if it is geometrically offset from the baryonic mass. The models using a $NFWSph$ with offset centres also failed to produce a detection.

\item \textbf{Can we constrain the inner dark matter slope?}

Table \ref{table:DarkMatter} shows that the spherical $gNFW$ model did not produce a Bayes factor above $20$ for any lens, again meaning this model could not be constrained using the lens data available. 

\end{itemize}

The results of model comparison demonstrate that dark matter profiles more complex than a $NFWSph$ cannot be constrained reliably using the lens imaging in this work. Nevertheless, in table \ref{table:DarkMatterParams} the inferred parameters of the dark matter distributions for these models are shown. The axis-ratios for the elliptical $NFW$ profiles are all around $0.95$ and therefore close to spherical. It is possible therefore that model comparison prefers a spherical $NFW$ simply because the inner dark matter of each lens is close to spherical. The inner slope of the $gNFW$ profiles tend to values below the $NFW$ value of $1$, but are consistent with it within errors, again raising the possibility that these Bayes factors are low because the true profiles are close to $1$. Given the low Bayes factors, we advise caution in the interpretation of these values and will seek to better understand how reliable our dark matter measurements are in future work.

Table \ref{table:DarkMatterParams} also shows the inferred parameters of which control the stellar mass distribution for models which assume a $NFWSph$, $NFWEll$ and $gNFWSph$ for the dark matter profile. For all three lenses, the inferred parameters governing the stellar mass profile ($\Psi_{\rm l}$, $R_{\rm l}$, $n_{\rm l}$, $I_{\rm l}$) are consistent with one another regardless of the assumed dark matter profile. The one exception is $\Psi_{l2}$ for SLACS3, which shows lower values for the $NFWEll$ (6.43) and $gNFWSph$ (6.78) models than the $NFWSph$ (7.71). These differences are small enough to have no significant impact on integrated quantities of the mass profile (e.g. masses, mass fractions, etc.).

We thus conclude that our lens models are sensitive to the inner dark matter halo masses, but cannot measure their geometry or inner density slope. Our inferred stellar mass profiles are insensitive to assumptions about dark matter. This was expected, given that the light profiles we infer for the lens galaxy are consistent for both the total-mass and decomposed mass profiles; if dark matter played an important role we would expect to see the decomposed model adjust its stellar mass distribution to compensate for the missing complexity in the dark matter component. Nevertheless, the lack of degeneracy between stellar and dark matter is somewhat surprising and will be investigated further in the future. We believe this is because of the differences between the stellar and dark matter distribution's density profiles and projected orientations.

\subsection{Masses and Surface Density Profiles}

\begin{figure*}
\centering
\includegraphics[width=0.33\textwidth]{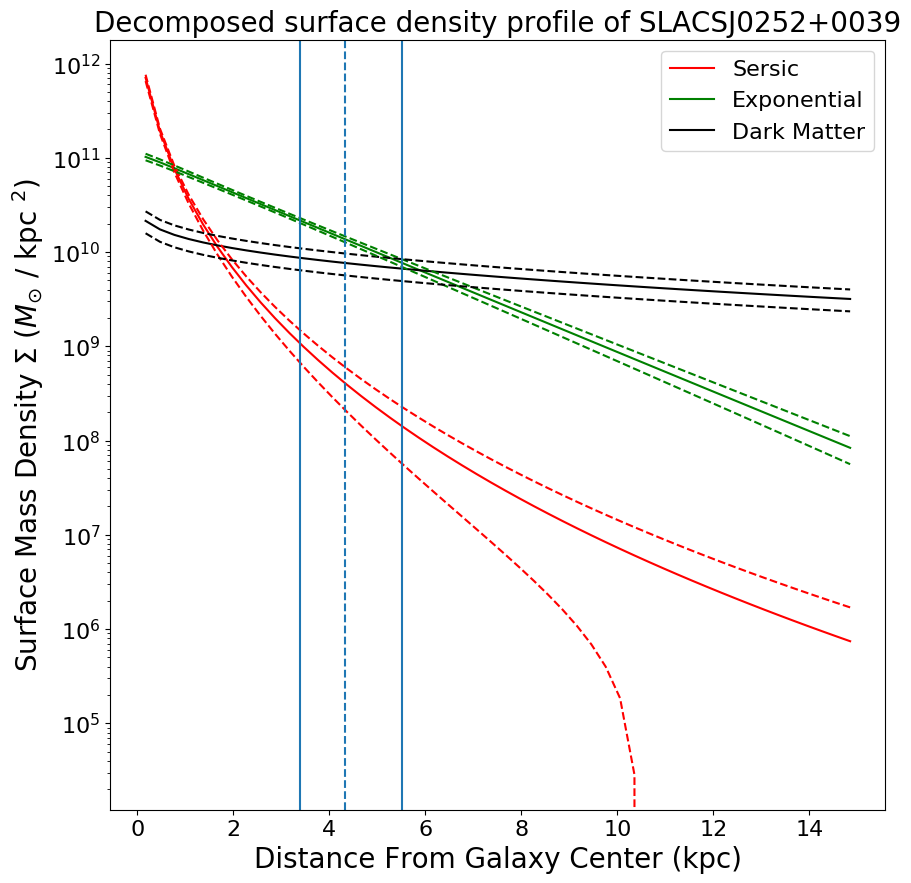}
\includegraphics[width=0.33\textwidth]{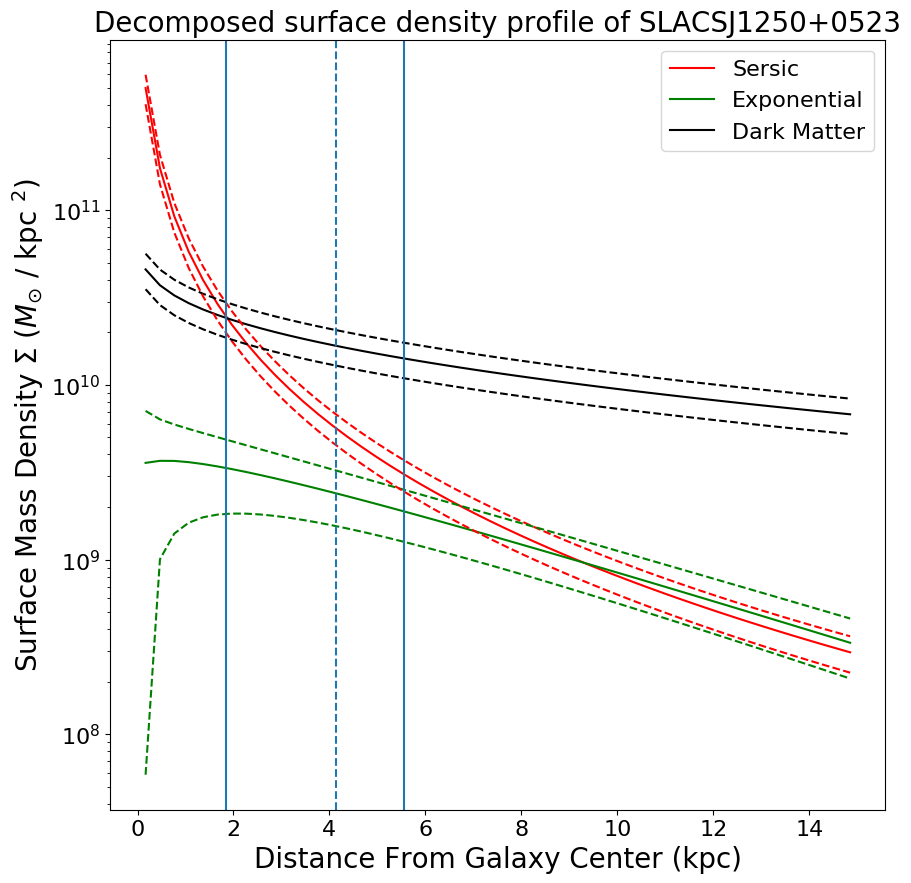}
\includegraphics[width=0.33\textwidth]{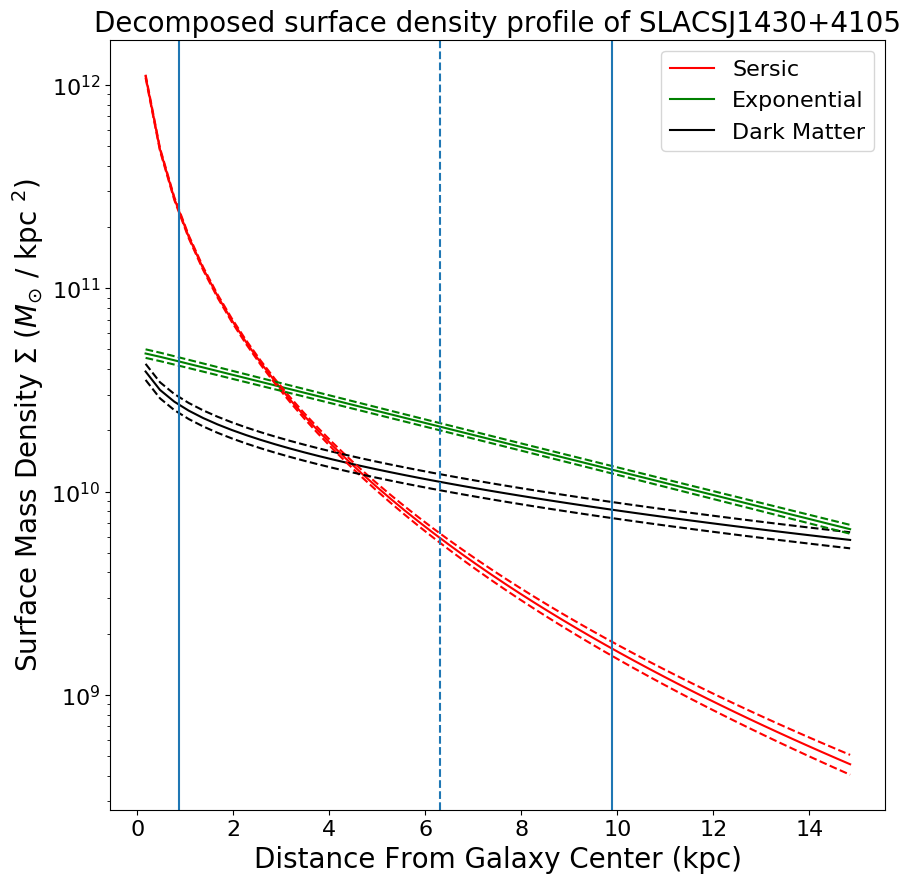}
\caption{One-dimensional surface density profiles of SLACS1 (left), SLACS2 (middle) and SLACS3 (right), showing the stellar mass density profile of the $Sersic$ component (red), $Exp$ component (green) and the $NFW$ dark matter profile (black). Densities are estimated using circular apertures centred on each lens galaxy. The dashed vertical blue lines show the Einstein radius of each lens and thick blue lines the radial extent of where the lensed source's light is visible. These indicate where the lensing data directly constrains the mass model. The density profile of the most probable model is given by the solid lines. To compute the models at $3\sigma$ errors, given by dashed lines, the density profiles of all models sampled by {\tt MultiNest} are weighted by their sampling probability and averaged so as to include the covariance between different models.}
\label{figure:Density}
\end{figure*}

\begin{table*}
\resizebox{\linewidth}{!}{
\begin{tabular}{ l | l | l | l | l | l | l } 
\multicolumn{1}{p{1.2cm}|}{Target Name} 
& \multicolumn{1}{p{1.5cm}|}{\centering Radius (kpc)} 
& \multicolumn{1}{p{1.6cm}|}{Total Mass ($10^{11} M_{\rm \odot}$)} 
& \multicolumn{1}{p{1.5cm}|}{Stellar Mass ($10^{11} M_{\rm \odot}$)} 
& \multicolumn{1}{p{1.5cm}|}{$Sersic$ Mass ($10^{11} M_{\rm \odot}$)}  
& \multicolumn{1}{p{1.5cm}|}{$Exponential$ Mass ($10^{11} M_{\rm \odot}$)}  
& \multicolumn{1}{p{1.5cm}|}{Dark Matter Mass ($10^{11} M_{\rm \odot}$)}
\\ \hline
& & & & & & \\[-4pt]
$\mathbf{SLACSJ0252+0039}$ & 10.0 & $3.12 \pm 0.18$ & $2.01 \pm 0.29$ & $0.43 \pm 0.05$ & $1.58 \pm 0.10$ & $1.11 \pm 0.30$ \\[2pt]
$\mathbf{SLACSJ0252+0039}$ & 500.0 & $33.53 \pm 8.08$ & $2.09 \pm 8.18$ & $0.43 \pm 0.05$ & $1.66 \pm 1.16$ & $31.14 \pm 8.21$ \\[2pt]
\hline
$\mathbf{SLACSJ1250+0523}$ & 10.0 & $4.78 \pm 0.03$ & $1.65 \pm 0.48$ & $1.127 \pm 0.25$ & $0.38 \pm 0.13$ & $3.12 \pm 0.72$ \\[2pt]
$\mathbf{SLACSJ1250+0523}$ & 500.0 & $91.12 \pm 20.00 $ & $2.36 \pm 20.00$  & $1.67 \pm 0.34 $ & $0.69 \pm 0.28 $ & $88.66 \pm 20.39$ \\[2pt]
\hline
$\mathbf{SLACSJ1430+4105}$ & 10.0 & $8.17 \pm 0.04$ & $6.19 \pm 0.16$ & $2.63 \pm 0.07$ & $3.56 \pm 0.15$ & $1.97 \pm 0.18$ \\[2pt]
$\mathbf{SLACSJ1430+4105}$ & 500.0 & $67.90 \pm 4.74 $ & $12.04 \pm 5.05$ & $2.98 \pm 0.11$ & $9.06 \pm 0.45$ & $55.96 \pm 5.07$ \\[2pt]
\end{tabular}
}
\caption{The mass measurements of each galaxy using the final model chosen by Bayesian model comparison, estimated within circular apertures of radii $10$ kpc and $500$ kpc. For each galaxy, the mass (column 3), stellar mass (column 4), $Sersic$ mass (column 5), $Exp$ mass (column 6) and dark matter mass (column 7) are given.}
\label{table:Masses}
\end{table*}

The highest Bayesian evidence model for the decomposed mass model fit to each lens galaxy is as follows:

\begin{itemize}

\item \textbf{SLACS1} - A $Sersic$ + $Exp$ + $NFWSph$ model, where the $Sersic$ and $Exp$ components are rotationally offset ($\approx 70^\circ$), centrally aligned and share the same mass-to-light ratio.

\item \textbf{SLACS2}- A $Sersic$ + $Exp$ + $NFWSph$ model, where the $Sersic$ and $Exp$ components are rotationally offset ($\approx 80^\circ$), centrally offset ($\approx  0.1$ kpc) and have independent mass-to-light ratios including a radial gradient in $Exp$ component.

\item \textbf{SLACS3} - A $Sersic$ + $Exp$ + $NFWSph$ model, where the $Sersic$ and $Exp$ components are rotationally offset ($\approx 20^\circ$), centrally offset ($\approx  0.3$ kpc) and have independent mass-to-light ratios.

\end{itemize}

Table \ref{table:Masses} gives mass estimates for each lens galaxy, using its highest Bayesian evidence decomposed model. The mass of the different components within radii of $10$ kpc and $500$ kpc, using circular apertures, are shown. Mass estimates are broadly consistent with those found in previous SLACS analysis, especially given the uncertainties that arise due to varying the initial mass function. Dark matter halo masses of order $10^{12-13}M_{\odot}$ are consistent with the expectation that we are studying massive elliptical galaxies. However, it should be noted that our lens model only constrains the inner dark matter halo profile directly. The extrapolation of this profile to large radii and thus mass estimates at $ \gtrsim 50$~kpc are dependent on what assumptions we make about their scale radii, which for this work were fixed to $10\,R_{\rm l}$. Our dark matter mass estimates at large radii may be subject to change if our assumptions about the scale radius are not reliable.

The surface density profiles of each lens's preferred model are displayed in figure \ref{figure:Density}. The $Sersic$ component is displayed in red, the $Exp$ component in green and the $NFW$ component in black. The vertical blue lines indicate the region in which light from the lensed source is observed (i.e. in which the lensing analysis directly constrains the mass profile). These plots are derived using circular apertures centered on each component of the mass model, therefore omitting any geometric offsets. For each lens, the first component dominates the inner ($ < 2$ kpc) regions, with the second component beginning to dominate around the Einstein radius at $5-10$ kpc. At larger radii ($> 15$ kpc), where our analysis becomes an extrapolation, the dark matter begins to dominate. Of the total stellar mass, the $Exp$ component is the dominant component in SLACS1 and SLACS3, contributing $80\%$ ($80\%$) and $57\%$ ($75\%$) within $10$ kpc ($500$ kpc) respectively. For SLACS2, it makes up just $35\%$ ($21\%$) within $10$ kpc ($500$ kpc). 

\subsection{The Role of Lensing}\label{LensingRole}

\begin{figure*}
\centering
\includegraphics[width=0.325\textwidth]{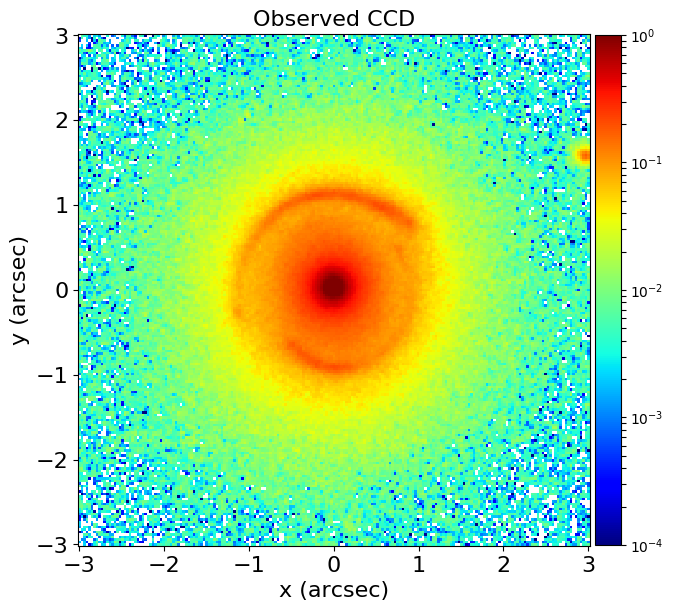}
\includegraphics[width=0.325\textwidth]{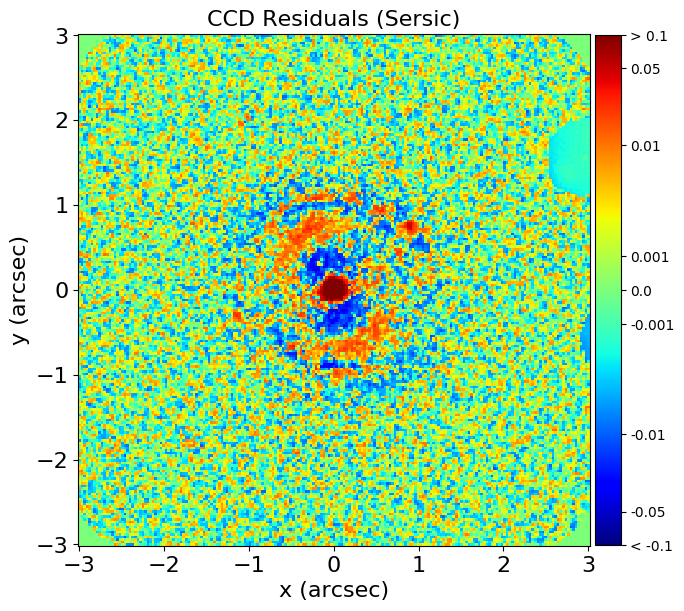}
\includegraphics[width=0.325\textwidth]{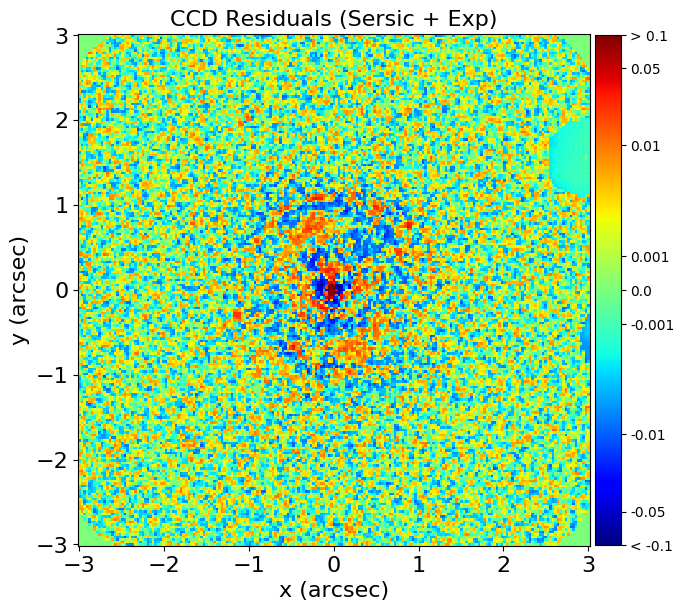}
\includegraphics[width=0.325\textwidth]{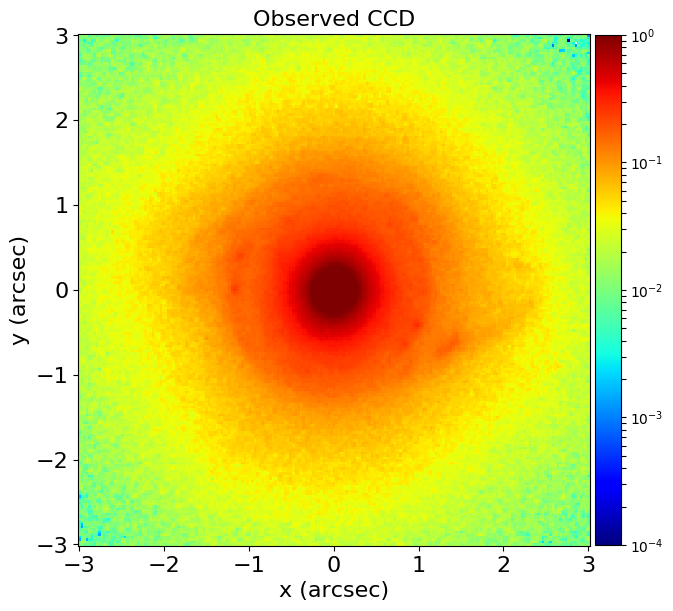}
\includegraphics[width=0.325\textwidth]{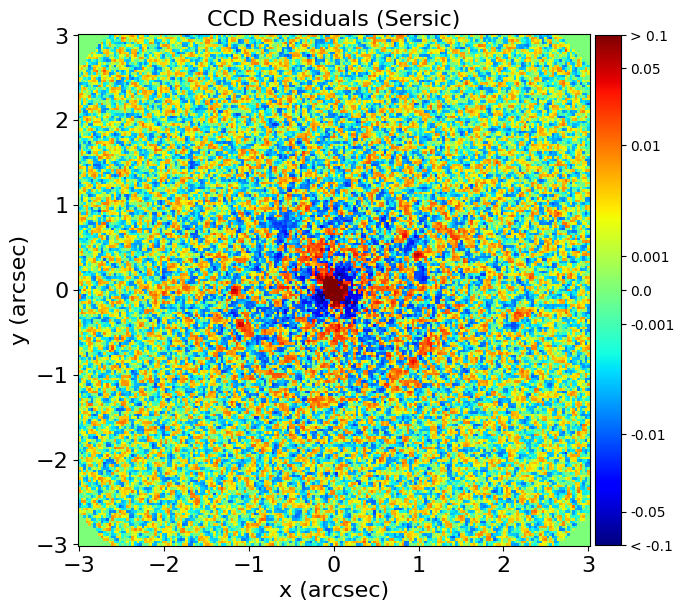}
\includegraphics[width=0.325\textwidth]{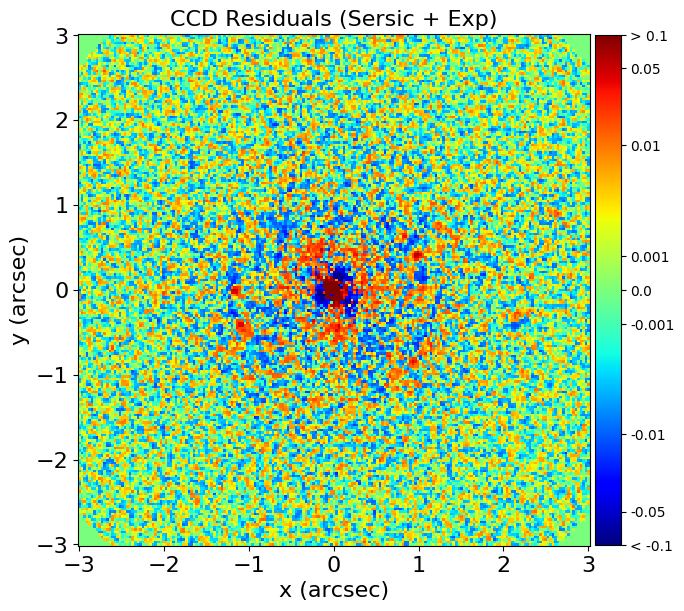}
\includegraphics[width=0.325\textwidth]{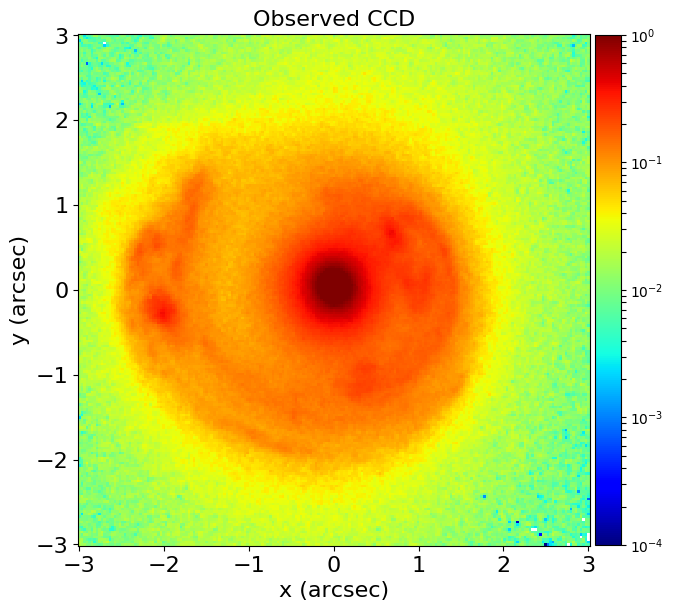}
\includegraphics[width=0.325\textwidth]{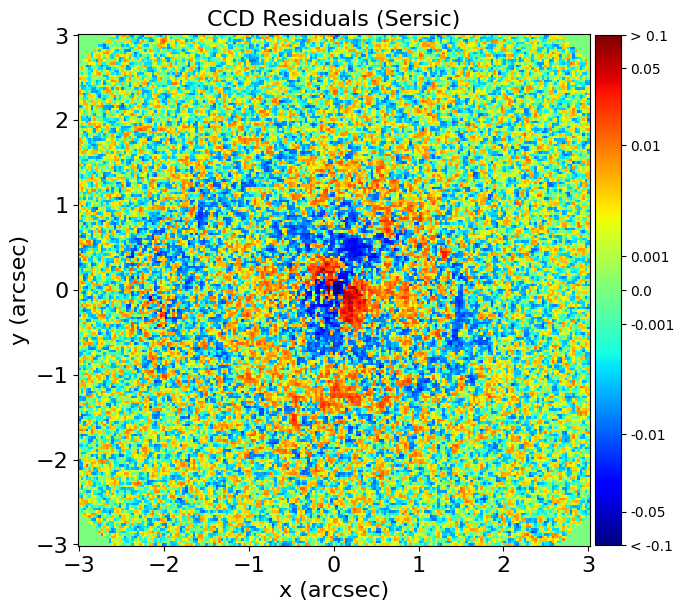}
\includegraphics[width=0.325\textwidth]{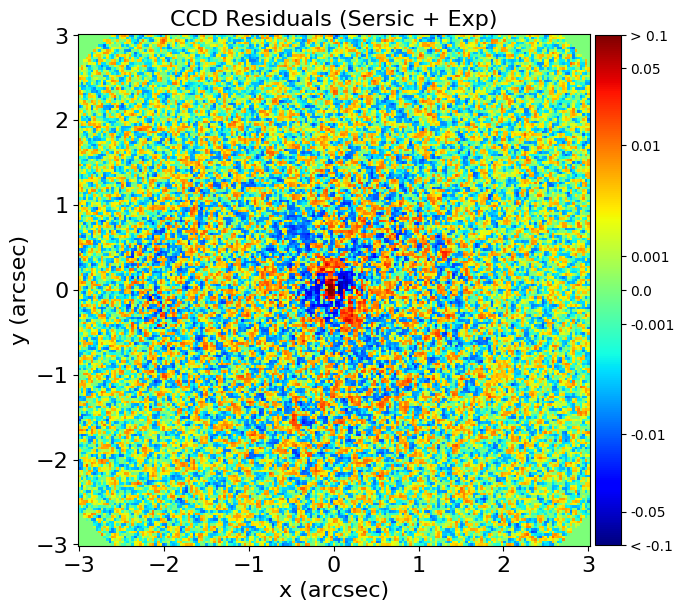}
\caption{A comparison of the residuals of the $Sersic$ and final chosen $Sersic$ + $Exp$ models using the decomposed mass-profile for all three images. The colourmap of the residuals is plotted using a symmetric log normalization and is trimmed to a small range in values to emphasise the difference in residuals between the two models. For all 3 lenses, the lens light fit can be seen to improve for the $Sersic$ + $Exp$ model, and in SLACS1 and SLACS3 the source residuals also improve.} 
\label{figure:LightResi}
\end{figure*}

Figure \ref{figure:LightResi} shows the residuals of the highest likelihood $Sersic$ model and chosen $Sersic$ + $Exp$ model, for the decomposed mass profiles. For all three lenses, the chosen two component models can be seen to improve the fit in the central regions of the lens galaxy, contributing to the increase in Bayesian evidence. However, in SLACS1 and SLACS3, improvements can also be seen visually in the residuals of the lensed source reconstruction, indicating that the strong lensing analysis is also contributing to the choice of the two-component model. Although improvements cannot be seen `by-eye' in SLACS2, we will now show they are also present in the fit.

We now seek a more quantitative measure of how much of this signal can be attributed to lensing. We define $\chi^2_{Src}$, the sum of $\chi^2$ values in pixels where the lensed source is located (source contribution values $> 0.2$, see N18), as a measure of the goodness-of-fit for the source. This comprises a different number of image pixels for the total mass and decomposed mass profile fits, therefore we renormalize each $\chi^2_{Src}$ value to that of the total-mass profile by multiplying by the ratio of number of pixels in each fit. We convert this to $\mathrm{ln}\,\epsilon_{Src} = -0.5\chi^2_{Src}$, the likelihood contribution of the source fit. The values of $\mathrm{ln}\,\epsilon_{Src}$ quoted correspond to the highest likelihood model at the end of each analysis, although there is little variation between models. 

{\tt PyAutoLens}'s Bayesian framework also measures the evidence of a model via a set of source-plane regularization terms, which quantify the complexity of the source reconstruction (such that worse mass models correspond to a more complex and lower evidence solution). $\chi^2$ is therefore not the only term which should be inspected to address how much the lensing analysis has improved. However, we find the regularization terms are driven by the quality of the lens subtraction, as opposed to the mass model, thus we omit them. Therefore, we define $\Delta\mathrm{ln}\,\epsilon_{Src} = \mathrm{ln}\,\epsilon_{Src,f} - \mathrm{ln}\,\epsilon_{Src,i}$ (i.e. final preferred model $-$ initial $Sersic$ model), to describe how much the lensing data contributes to the increase in evidence for a more complex model. Finally, we compare this value for the decomposed- and total-mass profiles, to ensure their different light subtractions do not impact our comparison.

For the total-mass profiles, we obtain $\Delta\mathrm{ln}\,\epsilon_{Src} = -7$ for SLACS1, $\Delta\mathrm{ln}\,\epsilon_{Src} = 9$ for SLACS2 and $\Delta\mathrm{ln}\,\epsilon_{Src} = 50$ for SLACS3. Since we use the same $SIE$ mass profile throughout, the source reconstruction only improves as a consequence of an improved lens subtraction leaving fewer residuals in the regions of the image the source is located, which is the case for SLACS3. For SLACS1 and SLACS2, the single $Sersic$ profile produced a sufficiently good lens subtraction over these regions. For the decomposed models, we obtain $\Delta\mathrm{ln}\,\epsilon_{Src} = 116$, $\Delta\mathrm{ln}\,\epsilon_{Src} = 61$ and $\Delta\mathrm{ln}\,\epsilon_{Src} = 152$ for SLACS1, SLACS2 and SLACS3 respectively. Thus, more complex light profiles are clearly necessary to better fit the lensing data.

\subsection{Physical Plausibility And Bayesian Evidence}

The Bayesian evidence is a powerful diagnostic in determining the lens model complexity. However, caution must be exercised in relying on just the increase in evidence as a means to determine the physical plausibility of a lens model. A model may fit the data better, and therefore give a large Bayes factor, but the actual reason for this may be a different scenario physically. Comparing different models, with different assumptions, is a useful way to test for this. For instance, offsets in the rotation angles and centroids of the $Sersic$ and $Exp$ components were seen for both mass model assumptions (total and decomposed), increasing our confidence that this is a physically genuine component of the lens model.

There are models where a large Bayes factor does not definitely rule out another physically plausible scenario. An example of this is our detection of a radial gradient in the stellar mass distribution. This model was chosen because the data required a more shallow mass distribution in the central regions of the lens galaxy of SLACS2, in a way our assumed dark matter profiles (including a $gNFW$) could not produce. It is plausible a more flexible dark matter model capable of removing more mass from the centre (e.g. a cored profile) would produce the same Bayes factor. It would be difficult to distinguish between these possibilities using just our lensing data and demonstrates a limitation of the method.   
\section{Discussion} \label{Discussion}

\subsection{The Formation and Structure of Massive Elliptical Galaxies}

We quantified the structure of three massive elliptical (ME) strong lenses using two different approaches. In the first approach we fit a light profile to each galaxy that was neither constrained by nor used to constrain the mass model of the same galaxy that we obtained from a strong lensing analysis. Two-component light profiles were preferred over single component light profiles with this approach for all three MEs, with Bayes factors of order $200-500$ (translating to a $\ge 10\sigma$ detection). The overall light profile of the one- and two-component fits were visibly similar, signifying that the improvement in the fit associated with the two-component model was spread over many thousands of image-pixels. Nevertheless, detailed inspection of the image residuals revealed that improvements from the two-component model could be seen in the fit to the lens galaxy.

The Sersic component of each lens was more compact (lower $R_l$), rounder (higher $q_l$) and more concentrated (higher $n_l$) than the Exponential component. Photometric decompositions of (non-lensing) MEs have inferred the same two-component (or more) structure of MEs \citep{Huang2013a, Oh2016}. They argue that this supports the forward models of ME formation (e.g. two-phase formation) outlined in the introduction, where the compact Sersic component is an old `red nugget' core of each ME formed in a rapid and dissipative event at high-redshift ($z > \sim 3$). The second component is a surrounding envelope of stellar material built via passive growth of mergers thereafter. However, with photometry alone, detecting this decomposition relies on an inflexion between the surface photometry of the two components being observable (e.g. \citet{Oh2016}) or a $\chi^2$ improvement in the fit (e.g. \citet{DSouza2014}). Due to projection effects it is not clear such an inflexion is always observable and $\chi^2$ thresholds are challenging to quantify robustly due to the potential of over-fitting.

By incorporating the lens's light profile into the mass model, such that it is simultaneously constrained by the lensing analysis, we are able to independently confirm whether these two components are physically distinct. Incorrect fits due to projection effects or over-fitting must lead to a degradation of the lensing analysis. For all three lenses, the two-component light model (now incorporated into the mass model) improved the lensing analysis compared to the single-component fit. Therefore, \textbf{lensing confirms that our three MEs are comprised of (at least) two physically distinct components which are consistent with forward models of ME formation (e.g. \citet{Cooper2013, Cooper2015}}).

As outlined in the introduction, the outer envelope that we model as an individual Exponential profile is anticipated to consist of many separate components, each corresponding to an accretion event in the MEs past. Therefore, what we model as an Exponential should be viewed as the superposition of these sub-components where its inferred properties represent some average of all of the (most significant) accretion events in the MEs history. Attempting to decompose this component further, into two or more components, is beyond the scope of this work, but could necessitate our lensing based approach, given that the blurring of these components in the photometric imaging likely makes them inseparable via just light profile fitting.

Our limited sample of just three objects means a more rigorous comparison between observations and theory is not yet warranted. We therefore turn our attention to what the inclusion of strong lensing provides and argue that, once we have analysed a sufficiently large sample, lensing will provide information vital to distinguishing models of ME formation.

\subsection{What Does Lensing Observe?}

Lensing reveals two crucial pieces of additional information about each ME that are not measured when using photometry alone. The first is the mass-to-light profile of each stellar component of the galaxy. This does not require stellar population models, circumventing the debated form of the initial mass function \citep{Treu2009} and could of course act an independent test of the IMF alongside studies such as \citet{Auger2009, Auger2009a} and \citet{Sonnenfeld2015}. In one lens, the two components had different mass-to-light ratios (SLACS3) and in another we observed a radial gradient (SLACS2 - see \citet{Sonnenfeld2018} for radial gradients in other SLACS lenses). The mass-to-light profiles for SLACS2 and SLACS3 removed mass from the central component relative to its outer envelope. This could indicate that internal baryonic processes (e.g. AGN feedback) can play a role in reshaping the stellar mass profile, or be a signature of super-massive black hole scouring \citep{Faber1997, Trujillo2004, Merritt2006}. Larger samples are necessary to investigate this further.

The second, and perhaps most crucial piece of information lensing measures, is the central mass of each MEs dark matter halo. Having a direct measurement of each galaxy's halo mass means that (using large samples) we can tie each galaxy (statistically) to its initial formation in the early Universe (e.g via halo merger trees). This can circumvent selections bias (e.g. progenitor bias) and offers a direct means by which to test ME formation in cosmological simulations. Combined with our inferences on the stellar density profile, we can distinguish what role a ME's environment (and thus merger history) plays in shaping its stellar distribution compared to internal baryonic processes.

\subsection{Structural Geometry}

The `outer envelope' of each ME was more elliptical than the central component ($q_{\mathrm{l}} = 0.65 - 0.8$) and rotationally offset in projection, with two of these offsets exceeding $70^\circ$. Isophotal twists are seen in other works \citep{Huang2013a, Oh2016}, but rarely at such large angles. This is likely because most studies use lower mass galaxies; larger rotational offsets are indeed seen by \citet{Goullaud2018} in ellipticals of similar masses to ours. These offsets are consistent with models of ME formation, provided that: (i) the outer envelope aligns with the direction of preferential merger accretion in its past (i.e. its surrounding environment) and; (ii) the central nugget's orientation can be decoupled from this direction during its formation. This decoupling is observed in the EAGLE simulations \citep{Velliscig2015} and explained by a highly energetic event at $z > 3$ (e.g. a major merger) decoupling the galaxies central baryonic structure from its surrounding environment, leaving the baryonic material accreted afterwards to trace it. More detailed theoretical studies are warranted, but large rotational offsets may hold an imprint of a ME's high-redshift formation, for example the merger ratio, direction and abundance of cold gas \citep{Hopkins2009}. 

In two lenses, the centre of the two stellar components were observed to be offset, by $\sim100$pc (SLACS2) and $\sim300$pc (SLACS3). Whilst these offsets were seen in the light profile fit, without lensing it would have been unclear whether or not they were due to dust in the central regions of the galaxy \citep{Goullaud2018}. These offsets are not two bimodal peaks of material in the central regions of the galaxy, but instead caused by lopsidedness in the stellar mass distribution at large radii (e.g. $> 10$kpc; \citet{Gomer2018}). Thus, the offsets are tracing how recently material was accreted onto the ME’s outskirts, given that material accreted more recently will not have had sufficient time to settle onto orbits centred on its overall potential well. Centroid offsets could in principle be used to estimate the time since a previous (major) merger, however the details are likely very complicated and depend on the merger in-fall velocities, directions and the galaxy's mass distribution. Unfortunately, we cannot be guided by simulations due to their softening lengths being above $300$pc (e.g. \citep{Schaller2015b}). 

\subsection{Implications for Strong Lensing}

Whilst our focus has been on showing how lensing can benefit photometric studies of galaxies, the reverse is also true. That is, for lensing systems in which the source does not provide a large amount of constraining information (e.g. it is only doubly-imaged, compact or pointlike), fitting the lens's light will better constrain its mass profile (given adequate assumptions about the mass-to-light ratio profile). Furthermore, the freedom of well known lensing degeneracies, like the mass-sheet transformation \citep{Falco1985, Schneider2013, Schneider2014b} and source-position transformation \citep{Schneider2014, Wertz2018} will be restricted, as the lens's stellar mass profile must coincide with the photometric data. These degeneracies also assume axisymmetry and may be alleviated by the geometric offsets detected in this work.

There are a number of techniques aiming to use strong lensing as a precise probe of cosmology \citep[e.g.\ ][]{Oguri2012, Suyu2013, Collett2014}. These methods require an accurate mass model for the lens. The mass profiles typically used in these studies do not capture the rotational and centroid offsets observed in this work \citep[although see][]{Birrer2018}, which have a unique impact on ray-tracing and could potentially impact the cosmological inference. Once we have analysed larger samples, we will investigate what impact, if any, these assumptions have on the inferred cosmology. Simultaneously fitting the lens's light and mass may also improve attempts to detect dark matter substructures via strong lensing (e.g. \citep{Vegetti2014}), where baryonic structures like an inclined disk mimic the substructure signal \citep{Hsueh2016, Hsueh2017, Hsueh2018}.

Investigating the `Fundamental Surface of Quads', \citet{Gomer2018} showed that the positions of quadruply imaged quasars provide evidence for rotational and centroid offsets in the mass distributions of lens galaxies, consistent with the models we have presented here. However, the authors only investigated offsets between the stellar and dark matter components in galaxies. Based on our work, we would suggest their inferences may instead be due to offsets between the different baryonic components of their lens galaxies. However, the selection function of strongly lensed quasars leads to a different lens galaxy population and further study is necessary to confirm this.

\subsection{Galaxy Structure With Strong Lensing}

Strong lensing promises to be a powerful technique to study galaxy structure. Euclid, LSST and other surveys are set to find of order 100,000 strong lenses in the next decade \citep{Collett2015}, spanning the entire Hubble diagram. Galaxies of later-type morphology will produce more distinct and thus measurable lensing effects than MEs, due to their distinct bulge-disk structure \citep{OrbandeXivry2008, Hsueh2017}. Whereas this study focused on each galaxy's two most dominant stellar components, future studies can investigate how sensitive lensing is to smaller galaxy structures (e.g. bars and psuedo-bulges) which are difficult to separate photometrically from the bulk of the galaxy's light. The light from structures that dominate well outside the Einstein radius of a lens, such as the stellar halo \citep{DSouza2014, Spavone2017}, will also be observable in future surveys and, in conjunction with weak lensing, will provide further constraints on the distribution of stellar and dark mass. 

Our technique complements currently ongoing large IFS surveys like the Mapping Nearby Galaxies at Apache Point Observatory \citep{Yan2016a}. These surveys are taking resolved spectra for $10000$ galaxies, enabling the comparison of their kinematic mass structure and photometry. However, these IFS surveys are restricted to the local Universe (e.g. $z < 0.02$), whereas lensing will span a broad range in lens galaxy redshifts (from $z \sim 0.05$ to $z > 3$). Thus, evolutionary studies of galaxy structure are only possible with lensing.



\section{Summary}\label{Summary}

We fitted three strong gravitational lenses with the lens modeling software {\tt PyAutoLens}. For each system, we compared two mass models: (i) a total-mass model, where the lens galaxy's structure is inferred independently from strong lensing and; (ii) a decomposed-mass model, where the lens galaxy's structure is folded into the lensing analysis and thus constrained by it. This allowed us to compare whether the structures we see in each lens galaxy's light translate to physically genuine mass structures that are necessary to fit the lensing data.

All three lens galaxies were massive ellipticals (ME). For each, we inferred that they are structurally composed of two components (a double $Sersic$ model) as opposed to one component (a single $Sersic$ model). The two-component model consists of a compact and round central nugget, surrounded by a flatter and more extended outer envelope. The two-component model was preferred, irrespective of whether the lensing constraints were used. However, including lensing increased the significance of each detection and thus confirmed these structures are physically genuine (as opposed due to over-fitting, dust or projection effects). Furthermore, the inclusion of lensing enabled us to measure two quantities that a photometric only analysis cannot: (i) the stellar mass distribution (without stellar population models) and; (ii) the inner dark matter halo mass of each galaxy.

We interpreted this two-component model in the context of forward models of ME galaxies \citep{Cooper2013, Cooper2015}. The central components are consistent with a central `red nugget' (e.g. \citealt{Trujillo2006, Oldham2016}) formed at high redshift ($z > 2$) due to a highly dissipative event, as evidenced by their structural parameters (high Sersic index, low effective radius). The outer components are extended envelopes of material (low Sersic index, high effective radius) accreted via mergers from redshift $2$ onwards. With just 3 objects, a rigorous test of ME formation was not warranted. However, by directly measuring dark matter masses, in the future we will be able to statistically tie a large sample of ME's to their counterparts in simulations and use their inferred stellar density profiles to decouple what role baryonic feedback processes and accretion play in their evolution. Thus, strong lensing provides a new test of ME formation, especially given recent theoretical works highlighting correlations between halo mass and stellar density profile (e.g. \citealt{Lovell2018, Wang2018}).

We investigated the geometry of these two components. We detect that they are rotationally offset from one another in projection, in two cases by over $70^{\circ}$. This suggests the dissipitive event which forms the central nugget changes its orientation relative to its surrounding local environment, which the outer envelope (being aligned with the direction of preferential accretion) traces. Such offsets are consistent with the EAGLE simulations \citep{Velliscig2015}. Centroid offsets between each component were also detected ($\sim 200$ kpc) and interpreted as a lopsidedness in the mass distribution of each galaxy \citep{Gomer2018}. This provides information on when the last episode of significant accretion took place, but may be challenging to extract because these offsets are below the softening length of modern cosmology simulations \citep{Schaller2015b}.

Over the next decade, surveys such as Euclid and LSST are set to amass samples in excess of 100,000 strong lenses, spanning the entire Hubble diagram from redshifts zero to three and beyond. The analysis demonstrated in this work offers a completely new view of a galaxy's structure, that is not limited to the ME galaxies investigated here. Direct measurements of a galaxy's host dark matter halo are crucial to understanding each galaxy's place in hierarchical structure formation, whilst their stellar density profiles provide insight into the role baryonic physics has in shaping each galaxy. In the future, we will seek to understand how lensing can further our understanding of the formation and evolution of galaxies of any Hubble type.

\section*{Acknowledgements}

We thank the anonymous referee whose suggestions improved the quality of the work. The authors thank Ian Smail, Russell Smith, Simon Dye and Andrew Robertson for helpful conversations.
JWN is supported by the UK Space Agency, through grant ST/N001494/1.
RJM is supported by a Royal Society University Research Fellowship.
APC is supported by the UK Science and Technology Facilities Council, through grant ST/P000541/1.

SLACS1 and SLACS3 were both observed in HST programme GO-10886 and SLACS2 in programme GO-10494.
This work used the DiRAC Data Centric system at Durham University, which is operated by the Institute for Computational Cosmology on behalf of the STFC DiRAC HPC Facility (\url{www.dirac.ac.uk}). This equipment was funded by BIS National e-Infrastructure capital grant ST/K00042X/1, STFC capital grant ST/H008519/1, and STFC DiRAC Operations grant ST/K003267/1 and Durham University. DiRAC is part of the UK National e-Infrastructure. 

\appendix
\section{Lens Models}\label{LensModels}

The lens light profile intensities $I$ and mass model deflection angles $\alpha_{\rm x,y}$ (via integration of the convergence $\kappa$) are computed using the elliptical coordinate system $\kappa (\vec{\xi})$, where $\xi = \sqrt{{x}^2 + y^2/q^2}$. Parameters associated with the lens's total mass profile have no subscript, light matter profile the subscript `l', dark matter profile the subscript `d' and external shear the subscript `sh'. The model centres, axis-ratios and rotation angles are, in most cases, treated as free parameters. Multiple component models are computed by summing each individual component’s intensities (for a light profile) or deflection angles (for a mass profile) and their parameters are given matching numerical subscripts (e.g. $x_{\rm 1}$, $x_{\rm l2}$, etc.). The light profile intensities are computed using an adaptive oversampling routine and mass model deflection angles an adaptive numerical integrator and bi-linear interpolation which are described in N18. The lens model changes throughout the analysis pipeline and section \ref{Results} describes how the final lens model is chosen.

The lens's light profile is modeled using one or more elliptical Sersic profiles
\begin{equation}
\label{eqn:Sersic}
I_{\rm  Ser} (\xi_{\rm l}) = I_{\rm  l} \exp \bigg\{ -k_{\rm l} \bigg[ \bigg( \frac{\xi_{\rm l}}{R_{\rm  l}} \bigg)^{\frac{1}{n_{\rm l}}} - 1 \bigg] \bigg\} ,
\end{equation}
which have up to seven free parameters: $(x_l,y_l)$, the light centre, $q_l$, the axis ratio, $\theta_l$, the orientation angle, $I_l$, the intensity at the effective radius $R_l$ and $n_l$, the Sersic index. $k_l$ is a function of $n_l$.  The exponential light profile $I_{Exp} (\xi_l)$ corresponds to $n_l = 1$.

For total-mass modeling a softened power-law ellipsoid ($SPLE$) density profile of form
\begin{equation}
\label{eqn:SPLEkap}
\kappa_{\rm  pl} (\xi) = \frac{(3 - \alpha)}{1 + q} \bigg( \frac{\theta_{\rm  E}}{\xi} \bigg)^{\alpha - 1} ,
\end{equation}
is used, where $\theta_E$ is the model Einstein radius in arc seconds. 
The power-law density slope is $\alpha$, and setting $\alpha = 2$ gives the singular isothermal ellipsoid (SIE) model. The inclusion of an external shear field is supported, which introduces two additional parameters, the shear magnitude $\gamma_{\rm  sh}$ and orientation of the semi-major axis measured counter-clockwise from north, $\theta_{\rm  sh}$.

For decomposed mass modeling, the Sersic profile given by equation \ref{eqn:Sersic} is used to give the light matter surface density profile
\begin{equation}
\label{eqn:Sersickap}
\kappa_{Ser} (\xi_l) = \Psi_l \bigg[\frac{q_{\rm l}  \xi_{\rm l}}{R_{\rm l}}\bigg]^{\Gamma_{\rm l}} I_{Ser} (\xi_l) \, \, ,
\end{equation}
where $\Psi_l$ gives the mass-to-light ratio in electrons per second and $\Gamma_{\rm l}$ can fold a radial dependence into the conversion of mass to light. The assumption used in N18 of a constant mass-to-light ratio is given for $\Gamma_{\rm l} = 0$ and models which assume this value or include $\Gamma_{\rm l}$ as a free parameter are investigated. If there are multiple light profile components these can either assume a single shared value of $\Psi_l$ or two independent values. The exponential convergence profile $\kappa_{Exp} (\xi_l)$ corresponds to $n_l = 1$. 

The dark matter component is given by a generalized Navarro-Frenk-White (NFW) profile, representing the universal density profile predicted for dark matter halos by cosmological N-body simulations \citep{Navarro1996, Zhao1996, Navarro1997} and with a volume mass density given by 
\begin{equation}
\label{eqn:NFWrho}
\rho = \frac{\rho_s}{(r/r_s)^\gamma (1 + r/r_s)^{3-\gamma}}.
\end{equation}
The halo normalization is given by $\rho_{\rm s}$ and $r_{\rm s}$, the scale radius, is fixed to the value $r_s = 10R_l$ \citep{Bullock2001} and the $NFW$ profile corresponds to the solution $\gamma_{\rm d} = 1$. Coordinates for the NFW profile are scaled by $r_s$, giving the scaled elliptical coordinate $\eta_d = \xi_d / r_s$. Deflection angles for spherical NFW models use the analytic solutions given by \citet{Golse2002} and N18, whereas elliptical models integrate using the formalism of \citep{Keeton2003b}.
\section{Lens Modeling Results}\label{LensResults}

The most probable lens models discussed in section \ref{Results} and their $3\sigma$ errors are given in tables \ref{table:SingleSersicParams}, \ref{table:TwoCompAlignParams}, \ref{table:TwoCompMisalignParams} and \ref{table:MassParams}.

\begin{table*}
\resizebox{\linewidth}{!}{
\begin{tabular}{ l | l | l | l | l | l | l } 
\multicolumn{1}{p{3.0cm}|}{\centering Target Name} 
& \multicolumn{1}{p{1.5cm}|}{\centering Mass \\ Profile} 
& \multicolumn{1}{p{1.4cm}|}{Mass-to-light Ratio} 
& \multicolumn{1}{p{0.7cm}|}{Parameters}
& \multicolumn{1}{p{0.7cm}}{} 
& \multicolumn{1}{p{0.7cm}}{} 
& \multicolumn{1}{p{0.7cm}}{}
\\ \hline
& & & & & & \\[-4pt]
$\mathbf{SLACSJ0252+0039}$ & Total & N/A & $R_{\mathrm{l}} = 1.92^{+0.57}_{-0.54}$ & $n_{\mathrm{l}} = 3.90^{+0.47}_{-0.47}$ & $q_{\mathrm{l}} = 0.940^{+0.028}_{-0.029}$ & $\theta_{\mathrm{l}} = 92^{+20}_{-20}$\\[2pt]
$\mathbf{SLACSJ0252+0039}$ & Decomposed & N/A &$ R_{\mathrm{l}} = 1.78^{+0.42}_{-0.28}$ & $n_{\mathrm{l}} = 3.60^{+0.40}_{-0.32}$ & $q_{\mathrm{l}} = 0.917^{+0.025}_{-0.021}$ & $\theta_{\mathrm{l}} = 94.4^{+8.9}_{-7.6}$ \\[2pt]
\hline
$\mathbf{SLACSJ1250+0523}$ & Total & N/A & $R_{\mathrm{l}} = 2.10^{+0.50}_{-0.50}$ & $n_{\mathrm{l}} = 5.31^{+0.49}_{-0.54}$ & $q_{\mathrm{l}} = 0.883^{+0.016}_{-0.017}$ & $\theta_{\mathrm{l}} = 71.2^{+5.1}_{-4.3}$ \\[2pt]
$\mathbf{SLACSJ1250+0523}$ & Decomposed & N/A & $R_{\mathrm{l}} = 2.11^{+0.61}_{-0.55}$ & $n_{\mathrm{l}} = 5.31^{+0.58}_{-0.52}$ & $q_{\mathrm{l}} = 0.882^{+0.020}_{-0.020}$ & $\theta_{\mathrm{l}} = 70.9^{+4.8}_{-4.4}$ \\[2pt]
\hline
 $\mathbf{SLACSJ1430+4105}$ & Total & N/A & $R_{\mathrm{l}} = 1.93^{+0.26}_{-0.27}$ & $n_{\mathrm{l}} = 4.73^{+0.28}_{-0.29}$ & $q_{\mathrm{l}} = 0.924^{+0.012}_{-0.011}$ & $\theta_{\mathrm{l}} = 87.0^{+4.5}_{-4.7}$ \\[2pt]
$\mathbf{SLACSJ1430+4105}$ & Decomposed & N/A & $R_{\mathrm{l}} = 2.41^{+0.29}_{-0.35}$ & $n_{\mathrm{l}} = 5.13^{+0.25}_{-0.30}$ & $q_{\mathrm{l}} = 0.9019^{+0.0084}_{-0.0078}$ & $\theta_{\mathrm{l}} = 85.6^{+2.4}_{-2.5}$\\[2pt]
\end{tabular}
}
\caption{The inferred most probable parameters (with $3\sigma$ errors) of the $Sersic$ profile, using a total mass $Sersic$ + $SIE$ model (rows 1, 3 and 5) and decomposed $Sersic$ + $NFWSph$ model (rows 2, 4 and 6). The inferred parameters of the $Sersic$ profile are consistent for both models, for all three lenses.}
\label{table:SingleSersicParams}
\end{table*}

\begin{table*}
\resizebox{\linewidth}{!}{
\begin{tabular}{ l | l | l | l | l | l | l } 
\multicolumn{1}{p{3.0cm}|}{\centering Target Name} 
& \multicolumn{1}{p{1.5cm}|}{\centering Mass \\ Profile} 
& \multicolumn{1}{p{1.4cm}|}{Mass-to-light Ratio} 
& \multicolumn{1}{p{0.7cm}|}{Parameters}
& \multicolumn{1}{p{0.7cm}}{} 
& \multicolumn{1}{p{0.7cm}}{} 
& \multicolumn{1}{p{0.7cm}}{}
\\ \hline
& & & & & & \\[-4pt]
$\mathbf{SLACSJ0252+0039}$ & Total & N/A & $R_{\mathrm{l1}} = 0.199^{+0.060}_{-0.046}$ & $n_{\mathrm{l1}} = 2.81^{+0.87}_{-0.76}$ & $q_{\mathrm{l1}} = 0.976^{+0.032}_{-0.043}$ & $\theta_{\mathrm{l1}} = 93^{+10}_{-10}$ \\[2pt]
 & & & $R_{\mathrm{l2}} = 1.01^{+0.12}_{-0.11}$ & $q_{\mathrm{l2}} = 0.863^{+0.060}_{-0.056}$ & & \\[2pt] 
$\mathbf{SLACSJ0252+0039}$ & Decomposed & Single & $R_{\mathrm{l1}} = 0.183^{+0.050}_{-0.045}$ & $n_{\mathrm{l1}} = 2.68^{+0.73}_{-0.70}$ & $q_{\mathrm{l1}} = 0.984^{+0.020}_{-0.026}$ & $\theta_{\mathrm{l1}} = 97.5^{+7.9}_{-7.9}$ \\[2pt]
 & & & $R_{\mathrm{l2}} = 0.96^{+0.11}_{-0.11}$ & $q_{\mathrm{l2}} = 0.841^{+0.035}_{-0.048}$  & & \\[2pt]
\hline
$\mathbf{SLACSJ1250+0523}$ & Total & N/A & $R_{\mathrm{l1}} = 0.60^{+0.18}_{-0.16}$ & $n_{\mathrm{l1}} = 3.61^{+0.57}_{-0.58}$ & $q_{\mathrm{l1}} = 0.875^{+0.017}_{-0.018}$ & $\theta_{\mathrm{l1}} = 71.0^{+4.2}_{-4.0}$ \\[2pt]
  & & &$R_{\mathrm{l2}} = 2.01^{+0.24}_{-0.21}$ & $q_{\mathrm{l2}} = 0.977^{+0.032}_{-0.047}$ & & \\[2pt]
$\mathbf{SLACSJ1250+0523}$ & Decomposed & Single & $R_{\mathrm{l1}} = 0.57^{+0.15}_{-0.14}$ & $n_{\mathrm{l1}} = 3.50^{+0.73}_{-0.71}$ & $q_{\mathrm{l1}} = 0.875^{+0.022}_{-0.024}$ & $\theta_{\mathrm{l1}} = 71.6^{+5.0}_{-4.9}$ \\[2pt]
 & & & $R_{\mathrm{l2}} = 2.02^{+0.30}_{-0.26}$ & $q_{\mathrm{l2}} = 0.978^{+0.031}_{-0.049}$ & & \\[2pt]
\hline
 $\mathbf{SLACSJ1430+4105}$ & Total & N/A &$R_{\mathrm{l1}} = 0.608^{+0.065}_{-0.063}$ & $n_{\mathrm{l1}} = 2.95^{+0.21}_{-0.21}$ & $q_{\mathrm{l1}} = 0.9329^{+0.0084}_{-0.0082}$ & $\theta_{\mathrm{l1}} = 81.5^{+2.5}_{-2.6}$ \\[2pt]
  & & & $R_{\mathrm{l2}} = 2.87^{+0.22}_{-0.22}$ & $q_{\mathrm{l2}} = 0.710^{+0.038}_{-0.043}$ & & \\[2pt]
 $\mathbf{SLACSJ1430+4105}$ & Decomposed & Single & $R_{\mathrm{l1}} = 0.571^{+0.055}_{-0.068}$ & $n_{\mathrm{l1}} = 2.727^{+0.19}_{-0.20}$ & $q_{\mathrm{l1}} = 0.928^{+0.007}_{-0.008}$ & $\theta_{\mathrm{l1}} = 81.8^{+2.4}_{-2.7}$ \\[2pt]
   & & & $R_{\mathrm{l2}} = 2.65^{+0.19}_{-0.24}$ & $q_{\mathrm{l2}} = 0.674^{+0.033}_{-0.041}$ & & \\[2pt]
\end{tabular}
}
\caption{The inferred most probable parameters (with $3\sigma$ errors) of the $Sersic$ + $Exp$ profiles with geometric parameters aligned between the two components. The results are for a total mass $Sersic$ + $Exp$ + $SIE$ model (rows 1, 3 and 5) and decomposed $Sersic$ + $Exp$ + $NFWSph$ model (rows 2, 4 and 6) where the mass-to-light ratio of the two components are the same. The inferred parameters of the $Sersic$ and $Exp$ profiles are consistent for both models, for all three lenses.}
\label{table:TwoCompAlignParams}
\end{table*}

\begin{table*}
\resizebox{\linewidth}{!}{
\begin{tabular}{ l | l | l | l | l | l | l } 
\multicolumn{1}{p{3.0cm}|}{\centering Target Name} 
& \multicolumn{1}{p{1.5cm}|}{\centering Mass \\ Profile} 
& \multicolumn{1}{p{1.4cm}|}{Mass-to-light Ratio} 
& \multicolumn{1}{p{0.7cm}|}{Parameters}
& \multicolumn{1}{p{0.7cm}}{} 
& \multicolumn{1}{p{0.7cm}}{} 
& \multicolumn{1}{p{0.7cm}}{}
\\ \hline
& & & & & & \\[-4pt]

$\mathbf{SLACSJ0252+0039}$ & Total & N/A & $R_{\mathrm{l1}} = 0.144^{+0.033}_{-0.028}$ & $n_{\mathrm{l1}} = 2.26^{+0.71}_{-0.65}$ & $q_{\mathrm{l1}} = 0.907^{+0.053}_{-0.053}$ & $\theta_{\mathrm{l1}} = 41^{+10}_{-20}$ \\[2pt]
 & & & $R_{\mathrm{l2}} = 0.879^{+0.11}_{-0.095}$ & $q_{\mathrm{l2}} = 0.830^{+0.049}_{-0.052}$ & $\theta_{\mathrm{l2}} = 105.0^{+9.2}_{-9.4}$ & $\theta = 99^{+10}_{-10}$ \\[2pt]
$\mathbf{SLACSJ0252+0039}$ & Decomposed & Single & $R_{\mathrm{l1}} = 0.141^{+0.036}_{-0.030}$ & $n_{\mathrm{l1}} = 2.36^{+0.76}_{-0.64}$ & $q_{\mathrm{l1}} = 0.889^{+0.059}_{-0.057}$ & $\theta_{\mathrm{l1}} = 38^{+20}_{-10}$ \\[2pt]
  & & & $R_{\mathrm{l2}} = 0.811^{+0.12}_{-0.093}$ & $q_{\mathrm{l2}} = 0.824^{+0.042}_{-0.044}$ & $\theta_{\mathrm{l2}} = 106.2^{+8.0}_{-8.3}$ & \\[2pt]
$\mathbf{SLACSJ0252+0039}$ & Decomposed & Independent & $R_{\mathrm{l1}} = 0.165^{+0.048}_{-0.042}$ & $n_{\mathrm{l1}} = 2.90^{+0.93}_{-0.87}$ & $q_{\mathrm{l1}} = 0.894^{+0.062}_{-0.070}$ & $\theta_{\mathrm{l1}} = 36^{+10}_{-10}$ \\[2pt]
& & & $R_{\mathrm{l2}} = 0.81^{+0.13}_{-0.11}$ & $q_{\mathrm{l2}} = 0.818^{+0.047}_{-0.051}$ & $\theta_{\mathrm{l2}} = 105.9^{+8.5}_{-8.4}$  & \\[2pt]
\hline
$\mathbf{SLACSJ1250+0523}$ & Total & N/A & $R_{\mathrm{l1}} = 0.99^{+0.29}_{-0.28}$ & $n_{\mathrm{l1}} = 4.48^{+0.60}_{-0.53}$ & $q_{\mathrm{l1}} = 0.882^{+0.020}_{-0.020}$ & $\theta_{\mathrm{l1}} = 70.9^{+4.7}_{-4.4}$ \\[2pt]
& & & $R_{\mathrm{l2}} = 2.11^{+0.31}_{-0.28}$ & $q_{\mathrm{l2}} = 0.75^{+0.18}_{-0.20}$ & $\theta_{\mathrm{l2}} = 163^{+18}_{-14}$ & $\theta = 104^{+20}_{-10}$ \\[2pt]
& & & $x_{\mathrm{l1}} = 0.0202^{+0.0028}_{-0.0024}$ & $y_{\mathrm{l1}} = 0.0098^{+0.0030}_{-0.0027}$ & $x_{\mathrm{l2}} = 0.049^{+0.025}_{-0.024}$ & $y_{\mathrm{l2}} = 0.022^{+0.027}_{-0.027}$ \\[2pt]
$\mathbf{SLACSJ1250+0523}$ & Decomposed & Single & $R_{\mathrm{l1}} = 0.61^{+0.23}_{-0.21}$ & $n_{\mathrm{l1}} = 3.62^{+0.50}_{-0.44}$ & $q_{\mathrm{l1}} = 0.874^{+0.019}_{-0.020}$ & $\theta_{\mathrm{l1}} = 72.0^{+4.3}_{-4.6}$ \\[2pt]
 & & & $R_{\mathrm{l2}} = 2.06^{+0.33}_{-0.24}$ & $q_{\mathrm{l2}} = 0.920^{+0.14}_{-0.15}$ & $\theta_{\mathrm{l2}} = 149^{+16}_{-17}$ \\[2pt]
 & & & $x_{\mathrm{l1}} = 0.021^{+0.0021}_{-0.0025}$ & $y_{\mathrm{l1}} = 0.010^{+0.0029}_{-0.0027}$ & $x_{\mathrm{l2}} =0.0361^{+0.0024}_{-0.0025}$ & $y_{\mathrm{l2}} = 0.0277^{+0.0031}_{-0.0028}$ \\[2pt]
 $\mathbf{SLACSJ1250+0523}$ & Decomposed & Independent & $R_{\mathrm{l1}} = 1.07^{+0.28}_{-0.26}$ & $n_{\mathrm{l1}} = 4.58^{+0.56}_{-0.62}$ & $q_{\mathrm{l1}} = 0.880^{+0.020}_{-0.018}$ & $\theta_{\mathrm{l1}} = 71.2^{+4.2}_{-4.4}$ \\[2pt]
 & & & $R_{\mathrm{l2}} = 2.11^{+0.44}_{-0.32}$ & $q_{\mathrm{l2}} = 0.70^{+0.11}_{-0.10}$ & $\theta_{\mathrm{l2}} = 162^{+16}_{-12}$ & \\[2pt]
 & & &$x_{\mathrm{l1}} = 0.0201^{+0.0025}_{-0.0024}$ & $y_{\mathrm{l1}} = 0.0091^{+0.0025}_{-0.0024}$ & $x_{\mathrm{l2}} = 0.056^{+0.028}_{-0.032}$ & $y_{\mathrm{l2}} = 0.036^{+0.022}_{-0.026}$ \\[2pt]
\hline
$\mathbf{SLACSJ1430+4105}$ & Total & N/A & $R_{\mathrm{l1}} = 0.571^{+0.058}_{-0.043}$ & $n_{\mathrm{l1}} = 2.873^{+0.612}_{-0.530}$ & $q_{\mathrm{l1}} = 0.924^{+0.021}_{-0.019}$ & $\theta_{\mathrm{l1}} = 93.0^{+4.3}_{-4.6}$ \\[2pt]
 & & & $R_{\mathrm{l2}} = 2.83^{+0.24}_{-0.25}$ & $q_{\mathrm{l2}} = 0.71^{+0.016}_{-0.021}$ & $\theta_{\mathrm{l2}} = 69.0^{+2.0}_{-1.8}$ & $\theta = 80.2^{+1.3}_{-0.9}$ \\[2pt]
 & & & $x_{\mathrm{l1}} = 0.028^{+0.0024}_{-0.0021}$ & $y_{\mathrm{l1}} = 0.031^{+0.0033}_{-0.0027}$ & $x_{\mathrm{l2}} = 0.003^{+0.021}_{-0.020}$ & $y_{\mathrm{l2}} = -0.045^{+0.021}_{-0.022}$ \\[2pt]
 $\mathbf{SLACSJ1430+4105}$ & Decomposed & Single & $R_{\mathrm{l1}} = 0.598^{+0.049}_{-0.063}$ & $n_{\mathrm{l1}} = 2.89^{+0.16}_{-0.18}$ & $q_{\mathrm{l1}} = 0.9214^{+0.0078}_{-0.0077}$ & $\theta_{\mathrm{l1}} = 92.8^{+2.5}_{-2.8}$  \\[2pt]
 & & & $R_{\mathrm{l2}} = 2.88^{+0.16}_{-0.20}$ & $q_{\mathrm{l2}} = 0.681^{+0.031}_{-0.028}$ & $\theta_{\mathrm{l2}} = 70.3^{+1.9}_{-2.4}$ & \\[2pt]
 & & & $x_{\mathrm{l1}} = 0.0283^{+0.0010}_{-0.0011}$ & $y_{\mathrm{l1}} = 0.0307^{+0.0010}_{-0.0011}$ & $x_{\mathrm{l2}} = -0.017^{+0.013}_{-0.016}$ & $y_{\mathrm{l2}} = -0.0095^{+0.014}_{-0.018}$ \\[2pt]
  $\mathbf{SLACSJ1430+4105}$ & Decomposed & Independent & $R_{\mathrm{l1}} = 0.587^{+0.042}_{-0.053}$ & $n_{\mathrm{l1}} = 2.86^{+0.14}_{-0.15}$ & $q_{\mathrm{l1}} = 0.922^{+0.0076}_{-0.0074}$ & $\theta_{\mathrm{l1}} = 92.0^{+2.2}_{-2.3}$  \\[2pt]
 & & & $R_{\mathrm{l2}} = 2.80^{+0.13}_{-0.17}$ & $q_{\mathrm{l2}} = 0.683^{+0.029}_{-0.029}$ & $\theta_{\mathrm{l2}} = 70.0^{+1.7}_{-2.2}$ & \\[2pt]
 & & & $x_{\mathrm{l1}} = 0.0280^{+0.0011}_{-0.0010}$ & $y_{\mathrm{l1}} = 0.0306^{+0.0010}_{-0.0011}$ & $x_{\mathrm{l2}} = -0.013^{+0.011}_{-0.015}$ & $y_{\mathrm{l2}} = -0.002^{+0.013}_{-0.017}$ \\[2pt]
\end{tabular}
}
\caption{The inferred most probable parameters (with $3\sigma$ errors) of the $Sersic$ + $Exp$ profiles with the alignment of geometric parameters corresponding to the highest evidence model chosen in the light profile geometry phase (table \ref{table:TwoCompOffset}). The results are for a total mass $Sersic$ + $Exp$ + $SIE$ model (rows 1, 4 and 7), decomposed $Sersic$ + $Exp$ + $NFWSph$ model (rows 2, 5 and 8) with shared mass-to-light ratio and independent mass-to-light ratios (rows 3, 6, and 9) are given.}
\label{table:TwoCompMisalignParams}
\end{table*}

\begin{table*}
\resizebox{\linewidth}{!}{
\begin{tabular}{ l | l | l | l | l | l | l } 
\multicolumn{1}{p{3.0cm}|}{\centering Target Name} 
& \multicolumn{1}{p{1.5cm}|}{\centering Mass \\ Profile} 
& \multicolumn{1}{p{1.4cm}|}{Mass-to-light Ratio} 
& \multicolumn{1}{p{0.7cm}|}{Parameters}
& \multicolumn{1}{p{0.7cm}}{} 
& \multicolumn{1}{p{0.7cm}}{} 
& \multicolumn{1}{p{0.7cm}}{}
\\ \hline
& & & & & & \\[-4pt]
$\mathbf{SLACSJ0252+0039}$ & Decomposed & Single & $\Psi_{\mathrm{l}} = 5.26^{+0.083}_{-0.076}$ & & & \\[2pt]
$\mathbf{SLACSJ0252+0039}$ & Decomposed & Independent & $\Psi_{\mathrm{l1}} = 4.17^{+0.368}_{-0.332}$ &  $\Psi_{\mathrm{l2}} = 5.64^{+0.231}_{-0.246}$ & & \\[2pt]
\hline
$\mathbf{SLACSJ1250+0523}$ & Decomposed & Single & $\Psi_{\mathrm{l}} = 1.27^{+0.062}_{-0.0085}$ & & & \\[2pt]
$\mathbf{SLACSJ1250+0523}$ & Decomposed & Independent &  $\Psi_{\mathrm{l1}} = 1.66^{+0.266}_{-0.245}$ &  $\Psi_{\mathrm{l2}} = 1.00^{+0.197}_{-0.187}$ & & \\[2pt]
$\mathbf{SLACSJ1250+0523}$ & Decomposed & Radial Gradient &  $\Psi_{\mathrm{l1}} = 1.56^{+0.301}_{-0.311}$ &  $\Psi_{\mathrm{l2}} = 1.20^{+0.247}_{-0.228}$ & $\Gamma_{\mathrm{l2}} = -0.24^{0.174}_{-0.154}$ & \\[2pt]
\hline
 $\mathbf{SLACSJ1430+4105}$ & Decomposed & Single & $\Psi_{\mathrm{l}} = 3.46^{+0.082}_{-0.096} $ & & & \\[2pt]
  $\mathbf{SLACSJ1430+4105}$ & Decomposed & Independent & $\Psi_{\mathrm{l1}} = 3.37^{+0.570}_{-0.69}$ &  $\Psi_{\mathrm{l2}} = 7.71^{+0.386}_{-0.355}$ & & \\[2pt]
\end{tabular}
}
\caption{The inferred most probable (with $3\sigma$ errors) inferred mass-to-light ratios $\Psi$ and radial gradient $\Gamma$ the $Sersic$ + $Exp$ profiles with the alignment of geometric parameters corresponding to the highest evidence model chosen in the light profile geometry phase (table \ref{table:TwoCompOffset}). The results are for decomposed $Sersic$ + $Exp$ + $NFWSph$ model.}
\label{table:MassParams}
\end{table*}

\begin{table*}
\resizebox{\linewidth}{!}{
\begin{tabular}{ l | l | l | l | l | l | l } 
\multicolumn{1}{p{3.0cm}|}{\centering Target Name} 
& \multicolumn{1}{p{1.5cm}|}{\centering Dark Matter \\ Profile} 
& \multicolumn{1}{p{0.7cm}|}{Parameters}
& \multicolumn{1}{p{0.7cm}}{} 
& \multicolumn{1}{p{0.7cm}}{} 
& \multicolumn{1}{p{0.7cm}}{}
\\ \hline
& & & & & & \\[-4pt]
$\mathbf{SLACSJ0252+0039}$ & $NFWSph$ & $\kappa_{\mathrm{d}} = 0.042^{+0.014}_{-0.011}$ & & & \\[2pt]
$\mathbf{SLACSJ0252+0039}$ &          &  $\Psi_{\mathrm{l}} = 5.26^{+0.083}_{-0.076}$ & $R_{\mathrm{l1}} = 0.141^{+0.036}_{-0.030}$ & $n_{\mathrm{l1}} = 2.36^{+0.76}_{-0.64}$  & $I_{\mathrm{l1}} = 0.448^{+0.125}_{-0.143}$ \\[2pt]
$\mathbf{SLACSJ0252+0039}$ &          &  & $R_{\mathrm{l2}} = 0.811^{+0.12}_{-0.093}$ & & $I_{\mathrm{l2}} = 0.075^{+0.021}_{-0.029}$ \\[2pt]
\hline
$\mathbf{SLACSJ0252+0039}$ & $NFWEll$ & $\kappa_{\mathrm{d}} = 0.044^{+0.016}_{-0.017}$ & $q_{\mathrm{d}} = 0.891^{+0.031}_{-0.034}$ & & \\[2pt]
$\mathbf{SLACSJ0252+0039}$ &          & $\Psi_{\mathrm{l}} = 5.11^{+0.475}_{-0.412}$ & $R_{\mathrm{l1}} = 0.186^{+0.065}_{-0.057}$ & $n_{\mathrm{l1}} = 3.07^{+0.98}_{-0.75}$ & $I_{\mathrm{l1}} = 0.282^{+0.062}_{-0.063}$ \\[2pt]
$\mathbf{SLACSJ0252+0039}$ &          &  & $R_{\mathrm{l2}} = 0.885^{+0.19}_{-0.12}$ &  & $I_{\mathrm{l2}} = 0.064^{+0.020}_{-0.024}$ \\[2pt]
\hline
$\mathbf{SLACSJ0252+0039}$ & $gNFWSph$ & $\kappa_{\mathrm{d}} = 0.041^{+0.025}_{-0.031}$ & $\gamma_{\mathrm{d}} = 0.688^{+0.542}_{-0.578}$ & & \\[2pt]
$\mathbf{SLACSJ0252+0039}$ &           & $\Psi_{\mathrm{l}} = 5.71^{+0.653}_{-0.736}$ &  $R_{\mathrm{l1}} = 0.166^{+0.042}_{-0.045}$ & $n_{\mathrm{l1}} = 2.83^{+0.78}_{-0.70}$ & $I_{\mathrm{l1}} = 0.333^{+0.072}_{-0.069}$ \\[2pt]
$\mathbf{SLACSJ0252+0039}$ &           &  &  $R_{\mathrm{l2}} = 0.820^{+0.22}_{-0.18}$ & & $I_{\mathrm{l2}} = 0.070^{+0.013}_{-0.009}$ \\[2pt]
\hline
$\mathbf{SLACSJ1250+0523}$ & $NFWSph$ & $\kappa_{\mathrm{d}} = 0.104^{+0.019}_{-0.020}$ & & & \\[2pt]
$\mathbf{SLACSJ1250+0523}$ &           & $\Psi_{\mathrm{l1}} = 1.66^{+0.266}_{-0.245}$ & $R_{\mathrm{l1}} = 1.07^{+0.28}_{-0.26}$ & $n_{\mathrm{l1}} = 4.58^{+0.56}_{-0.62}$ & $I_{\mathrm{l1}} = 0.106^{+0.014}_{-0.016}$ \\[2pt]
$\mathbf{SLACSJ1250+0523}$ &           & $\Psi_{\mathrm{l2}} = 1.00^{+0.197}_{-0.187}$ & $R_{\mathrm{l2}} = 2.11^{+0.44}_{-0.32}$ &  & $I_{\mathrm{l2}} = 0.025^{+0.008}_{-0.07}$ \\[2pt]
\hline
$\mathbf{SLACSJ1250+0523}$ & $NFWEll$ & $\kappa_{\mathrm{d}} = 0.090^{+0.022}_{-0.021}$ & $q_{\mathrm{d}} = 0.976^{+0.021}_{-0.021}$ & & $I_{\mathrm{l2}} = 0.0^{+0.0}_{-0.0}$ \\[2pt]
$\mathbf{SLACSJ1250+0523}$ &          & $\Psi_{\mathrm{l1}} = 1.72^{+0.376}_{-0.311}$  & $R_{\mathrm{l1}} = 0.91^{+0.42}_{-0.39}$ & $n_{\mathrm{l1}} = 4.24^{+0.76}_{-0.65}$ & $I_{\mathrm{l1}} = 0.115^{+0.013}_{-0.017}$ \\[2pt]
$\mathbf{SLACSJ1250+0523}$ &           & $\Psi_{\mathrm{l2}} = 1.11^{+0.326}_{-0.276}$ & $R_{\mathrm{l2}} = 2.17^{+0.49}_{-0.33}$ &  & $I_{\mathrm{l2}} = 0.025^{+0.007}_{-0.008}$ \\[2pt]
\hline
$\mathbf{SLACSJ1250+0523}$ & $gNFWSph$ & $\kappa_{\mathrm{d}} = 0.105^{+0.033}_{-0.031}$ & $\gamma_{\mathrm{d}} = 0.846^{+0.540}_{-0.631}$ & & \\[2pt]
$\mathbf{SLACSJ1250+0523}$ &           & $\Psi_{\mathrm{l1}} = 1.86^{+0.322}_{-0.365}$ &  $R_{\mathrm{l1}} = 0.91^{+0.44}_{-0.40}$ & $n_{\mathrm{l1}} = 4.28^{+0.71}_{-0.66}$ & $I_{\mathrm{l1}} = 0.114^{+0.016}_{-0.011}$ \\[2pt]
$\mathbf{SLACSJ1250+0523}$ &           & $\Psi_{\mathrm{l2}} = 0.97^{+0.254}_{-0.289}$ & $R_{\mathrm{l2}} = 2.21^{+0.48}_{-0.39}$ &  & $I_{\mathrm{l2}} = 0.025^{+0.009}_{-0.008}$ \\[2pt]
\hline
$\mathbf{SLACSJ1430+4105}$ & $NFWSph$ & $\kappa_{\mathrm{d}} = 0.054^{+0.020}_{-0.019}$ & & & \\[2pt]
$\mathbf{SLACSJ1430+4105}$ &          & $\Psi_{\mathrm{l1}} = 3.37^{+0.570}_{-0.691}$  & $R_{\mathrm{l1}} = 0.587^{+0.042}_{-0.053}$ & $n_{\mathrm{l1}} = 2.86^{+0.14}_{-0.15}$ & $I_{\mathrm{l1}} = 0.166^{+0.053}_{-0.047}$ \\[2pt]
$\mathbf{SLACSJ1430+4105}$ &          & $\Psi_{\mathrm{l2}} = 7.71^{+0.386}_{-0.355}$ & $R_{\mathrm{l2}} = 2.80^{+0.13}_{-0.17}$ &  & $I_{\mathrm{l2}} = 0.016^{+0.006}_{-0.006}$ \\[2pt]
 \hline
$\mathbf{SLACSJ1430+4105}$ & $NFWEll$ & $\kappa_{\mathrm{d}} = 0.077^{+0.025}_{-0.026}$ & $q_{\mathrm{d}} = 0.960^{+0.021}_{-0.024}$ & & \\[2pt]
$\mathbf{SLACSJ1430+4105}$ &          & $\Psi_{\mathrm{l1}} = 3.30^{+0.612}_{-0.710}$  & $R_{\mathrm{l1}} = 0.632^{+0.051}_{-0.054}$ & $n_{\mathrm{l1}} = 2.93^{+0.17}_{-0.15}$ & $I_{\mathrm{l1}} = 0.153^{+0.038}_{-0.042}$  \\[2pt]
$\mathbf{SLACSJ1430+4105}$ &          & $\Psi_{\mathrm{l2}} = 6.43^{+0.431}_{-0.452}$ & $R_{\mathrm{l2}} = 3.04^{+0.19}_{-0.22}$ &  & $I_{\mathrm{l2}} = 0.014^{+0.005}_{-0.006}$ \\[2pt]
\hline
$\mathbf{SLACSJ1430+4105}$ & $gNFWSph$ & $\kappa_{\mathrm{d}} = 0.099^{+0.036}_{-0.042}$ & $\gamma_{\mathrm{d}} = 0.774^{+0.431}_{-0.545}$ & & \\[2pt]
$\mathbf{SLACSJ1430+4105}$ &           & $\Psi_{\mathrm{l1}} = 3.31^{+0.742}_{-0.821}$ & $R_{\mathrm{l1}} = 0.59^{+0.061}_{-0.059}$ & $n_{\mathrm{l1}} = 2.84^{+0.17}_{-0.18}$ & $I_{\mathrm{l1}} = 0.170^{+0.042}_{-0.041}$ \\[2pt]
 $\mathbf{SLACSJ1430+4105}$ &          & $\Psi_{\mathrm{l2}} = 6.78^{+0.568}_{-0.611}$ & $R_{\mathrm{l2}} = 2.93^{+0.23}_{-0.24}$ &  & $I_{\mathrm{l2}} = 0.015^{+0.005}_{-0.007}$ \\[2pt]
\end{tabular}
}
\caption{The inferred most probable (with $3\sigma$ errors) inferred dark matter lensing masses $\kappa_{\mathrm{d}}$, dark matter axis ratios $q_{\mathrm{d}}$, dark matter inner density slopes $\gamma_{\mathrm{d}}$, mass-to-light ratios $\Psi$, effective radii $R_{\mathrm{l}}$, Sersic indexes $n_{\mathrm{l}}$ and intensities $I_{\mathrm{l}}$. These models use the $Sersic$ + $Exp$ profiles with the alignment of geometric parameters corresponding to the highest evidence model chosen in the light profile geometry phase (table \ref{table:TwoCompOffset}). The results are shown for three dark matter profiles, a spherical NFW $NFWSph$, an elliptical NFW $NFWEll$ and spherical generalized NFW $gNFWSph$.}
\label{table:DarkMatterParams}
\end{table*}

\bibliography{library}            

\label{lastpage}

\end{document}